\documentclass[12pt]{article}
\pdfoutput=1

\usepackage{amsmath}
\usepackage{amsfonts}
\usepackage{amssymb}

\usepackage[
      colorlinks=true,
      linkcolor=blue,
      urlcolor=blue,
      filecolor=black,
      citecolor=red,
      pdfstartview=FitV,
      pdftitle={},
        pdfauthor={ Michael Gutperle, John D. Miller},
        pdfsubject={},
        pdfkeywords={},
        pdfpagemode={},
        bookmarksopen=true
      ]{hyperref}
\hypersetup{linktocpage}

\usepackage{color}
\usepackage{graphicx}

\usepackage{sectsty}
\sectionfont{\large}

\renewcommand{\thefootnote}{\fnsymbol{footnote}}
\renewcommand{\thanks}[1]{\footnote{#1}}
\newcommand{\starttext}{
\setcounter{footnote}{0}
\renewcommand{\thefootnote}{\arabic{footnote}}}
\newcommand{\bea}{\begin{eqnarray}}
\newcommand{\eea}{\end{eqnarray}}
\newcommand{\be}{\begin{equation}}
\newcommand{\ee}{\end{equation}}

\numberwithin{equation}{section}

\def\det{{\rm det}}

\usepackage{ dsfont } %for identity matrix

\newcommand{\paren}[1]{\left(#1\right)}

\long\def\symbolfootnote[#1]#2{\begingroup%
\def\thefootnote{\fnsymbol{footnote}}\footnote[#1]{#2}\endgroup}

\begin{document}
\setlength{\baselineskip}{18pt}

\starttext
\setcounter{footnote}{0}

\begin{flushright}
\today
\end{flushright}

\bigskip

\begin{center}

{\Large \bf   Entanglement entropy at CFT junctions}

\vskip 0.4in

{\large   Michael Gutperle and John D. Miller}

\vskip .2in

{ \it Mani L. Bhaumik Institute for Theoretical Physics }\\
{ \it Department of Physics and Astronomy }\\
{\it University of California, Los Angeles, CA 90095, USA}\\[0.5cm]

\bigskip
\href{mailto:gutperle@physics.ucla.edu}{\texttt{gutperle}}\texttt{, }
\href{mailto:johnmiller@physics.ucla.edu}{\texttt{johnmiller@physics.ucla.edu}}

\bigskip

\bigskip

\end{center}
 
\begin{abstract}

\setlength{\baselineskip}{18pt}

We consider entanglement through permeable junctions of $N$ $(1+1)$-dimensional free boson and free fermion conformal field theories. In the folded picture we constrain the form of the general boundary state. We calculate replicated partition functions with interface operators inserted in the partially-folded picture, from which the entanglement entropy is calculated. The functional form of the universal and constant terms are the same as the $N=2$ case, depending only of the total transmission of the junction and the unit volume of the zero mode lattice. For $N>2$ we see a sub-leading divergent term which does not depend on the parameters of the junction. For $N=3$ we consider some specific geometries and discuss various limits.
\end{abstract}

\setcounter{equation}{0}
\setcounter{footnote}{0}

\newpage

\tableofcontents

\newpage

\section{Introduction}
\label{sec1}

The entanglement entropy of a  region ${\cal A}$  is given by the von Neumann entropy of the reduced  density matrix produced by tracing over all degrees of freedom in the complement $\bar{\cal A}$. This quantity provides a measure of the entanglement of the two regions and has been utilized in a wide variety of areas ranging from quantum information theory to black hole physics. 

In the present paper we are interested in entanglement entropy in two dimensional conformal field theories, studied first in   \cite{Holzhey:1994we,Calabrese:2004eu}.
For a spatial region given by a finite interval of length $L$ and  an UV cutoff $\epsilon$, the entanglement entropy of this interval has the following form 
\begin{equation}\label{basicEE}
\mathcal{S}_L=\frac{c}{3}\log\frac{L}{\epsilon}+C
\end{equation}
The logarithmically divergent term is universal and only depends on the central charge $c$  of the CFT. On the other hand the constant $C$ is in general regulator dependent and not universal. 

For a CFT with a boundary, defect or interface it was argued in  \cite{Calabrese:2004eu,Azeyanagi:2007qj}  that  the constant term $C$  becomes physically meaningful and is closely related to the boundary entropy first introduced in \cite{Affleck:1991tk}.
In this paper we will only consider conformal interfaces and there are two cases which one can distinguish. 

First, we  consider an interval placed symmetrically across an interface $\mathcal{I}$ between two CFTs of the same central charge, whose entanglement entropy is
\begin{equation}\label{symEE}
\mathcal{S}_\text{sym}=\frac{c}{3}\log\frac{L}{\epsilon}+C'(\mathcal{I})
\end{equation}
where now the constant term $C'$ is a function of the parameters of the interface $\mathcal{I}$. The universal term has the same form between (\ref{basicEE}) and (\ref{symEE}) as the endpoints of the interval where entanglement is strongest are symmetrically positioned away from the location of the interface. 

Second,   we  can locate the interface at the boundary of the region ${\cal A}$ and enlarge it to cover the whole of one of the two CFTs in the limit as $L$ becomes very large, so that the end-point of the interval is fixed to the location of the interface. It was shown in \cite{Sakai:2008tt} that  the central charge $c$ for universal term gets replaced by   a function of the parameters of the interface
\begin{equation}\label{asym2EE}
\mathcal{S}_\text{asym}=\frac{c}{3}\,f(\mathcal{I})\log\frac{L}{\epsilon}+\tilde{C}(\mathcal{I})
\end{equation}
The function $f(\mathcal{I})$ varies depending on the CFT and is known only for a few cases, two of which are reviewed in section \ref{sec3}. However, in general $f(\mathcal{I})$ must obey some limits. For an interface that completely decouples the two CFTs it must be the case that $f(\mathcal{I})=0$, while for an interface that completely transmits energy (so-called topological interfaces) it must be the case that $f(\mathcal{I})=1/2$\footnote{The reason that $f(\mathcal{I})=1/2$ instead of $1$  has to do with the fact that we are now considering an semi-infinite entangling interval with only one end-point, and thus should have half the entropy of the two end-point case in (\ref{basicEE}).}.

A natural generalization of an interface $\mathcal{I}$ connecting two CFTs is a junction $\mathcal{J}$ connecting $N$ CFTs along a common line. If we consider an entangling region containing one of the CFTs, say CFT$_i$, then the entanglement entropy has the same generic form as (\ref{asym2EE}); that is,
\begin{equation}\label{asymNEE}
\mathcal{S}_i=\frac{c}{3}\,f_{N,i}(\mathcal{J})\log\frac{L}{\epsilon}+\tilde{C}_{N,i}(\mathcal{J})
\end{equation}
For junctions between non-relativistic theories, it was shown in \cite{Calabrese:2011ru} that the universal term of (\ref{asymNEE}) is related to the universal term of (\ref{asym2EE}) via 
\begin{equation}\label{Calacon}
f_{N,i}(\mathcal{J})=f\big(\sqrt{\mathcal{T}_i}\,\big)
\end{equation}
where $\mathcal{T}_i$ is the total transmission coefficient from $i$-th theory to the other theories in the junction, however this has not been shown to hold in the conformal setting. In this work we will show that this relationship holds for arbitrary junctions between CFTs which are constructed from  free conformal bosons and fermions.

The paper is organized as follows. In section \ref{sec2} we review the folding trick  which turns the problem of constructing conformal interfaces into one of constructing boundary states. We review the construction of bosonic as well as fermionic boundary states and determine the normalization using the Cardy condition.
 In section \ref{sec3} we review the calculation of entanglement entropy  in the presence of a  bosonic and fermionic interface which is located at the boundary of the entangling space. In section \ref{sec4} we calculate the entanglement entropy of bosonic and fermionic $N$-junctions, generalizing the method introduced in the previous section. In section \ref{sec5} we construct all boundary states corresponding to $3$-junctions and discuss various features and limits. In section \ref{sec6} we summarize the main results of our work and discuss possible avenues for future work involving CFT junctions. Our conventions for the free boson and fermion CFTs, special functions as well as calculational details involving Gaussian integrals and circular determinants are relegated to appendices.

\section{CFT construction of interfaces and junctions }
\label{sec2}

\begin{figure}
\centering
\includegraphics[scale=1.0]{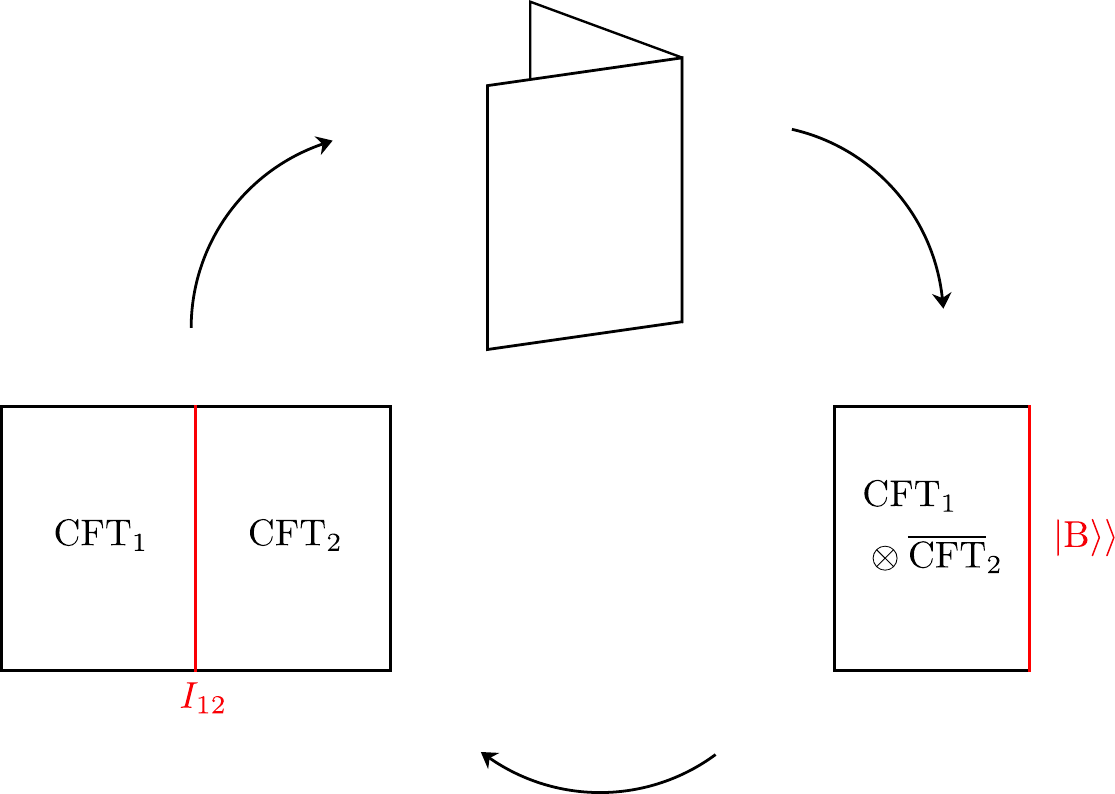}
\caption{Illustration of the parity transformation relating the interface between CFT$_1$ and CFT$_2$ to the tensor product CFT$_1\otimes\overline{\text{CFT}}_2$ with boundary. The folded picture is useful for characterizing classes of interfaces and some simple calculations (see \cite{Bachas:2001vj}). For calculations such as ours the boundary states need to be unfolded once they are found.\label{intfoldfig}}
\end{figure}

A conformally invariant interface between general CFT$_1$ and CFT$_2$ is described by an operator located at the interface that satisfies
\begin{equation}\label{intcond}
\paren{L^1_n-\bar{L}^1_{-n}}I_{12}=I_{12}\paren{L^2_n-\bar{L}^2_{-n}}
\end{equation}
for $n\geq 0$, where $L^i_n$ and $\bar{L}^i_n$ with $i=1,2$ are the Virasoro generators of each CFT. Finding operators that satisfy (\ref{intcond}) can be mapped to finding conformal boundary states satisfying
\begin{equation}\label{Virasorobndcond}
\paren{L^\text{total}_n-\bar{L}^\text{total}_{-n}}|\text{B}\rangle\rangle=0
\end{equation}
by use of a parity transformation. This is the content of the folding trick \cite{Bachas:2001vj}, which is illustrated in figure \ref{intfoldfig}. For general CFTs, the boundary states satisfying (\ref{Virasorobndcond}) are often difficult to find. When $c_\text{total}<1$  the CFT are rational an for a finite number of primary fields all solutions to (\ref{Virasorobndcond}) have been found \cite{Ishibashi:1988kg} and organized into modular invariant boundary states \cite{CARDY1989581}. However, since we are considering an $N$-times tensor product CFT in the folded picture ($N=2$ for interfaces, $N>2$ for junctions), the resulting folded CFT always has $c>1$ and  hence not rational. 
If one imposes additional conditions such as preservation of a current algebra or permutation symmetry,  more general constructions of boundary states and  interfaces are possible \cite{Bachas:2012bj,Recknagel:2002qq,Brunner:2005fv}.
 Another possibility is given by strengthening  the conditions (\ref{Virasorobndcond}) to boundary states satisfying
\begin{equation}
\paren{L^i_n-\bar{L}^i_{-n}}|\text{B}\rangle\rangle=0
\end{equation}
for each $i=1,2$ separately. This leads to so called topological defects or interfaces \cite{Bachas:2007td,Fuchs:2007tx,Brunner:2013ota}.
In this case solutions are known for wider classes of CFTs; e.g. for topological interfaces in rational CFTs
 the corresponding interface operators were found in \cite{Gutperle:2015kmw} and \cite{Brehm:2015plf} by building off of the modular invariant projection operators constructed in \cite{Petkova:2000ip}. When considering free fields, as in this work, the conditions can be written in terms of the creation and annihilation operators and can be solved by a coherent state anzatz. We will now show how this works for free bosonic interfaces and junctions.

\subsection{Bosonic interfaces}

Under the replacement $a^i_{n}\rightarrow S_{ij}\,\bar{a}^j_{-n}$ for a $2\times 2$ matrix $S$, the operator combinations in the generators $L_n^i$ are altered as
\begin{equation}\label{opreplace}
:a^i_{n-m}a^i_m:\,\,\longrightarrow\,\,S_{ij}\,S_{ik}:\bar{a}^j_{m-n}\bar{a}^k_{-m}:
\end{equation}
Considering summation over the index $i$ in the above and form of the generators (\ref{bosenVir}), it is seen that $L^\text{total}_n\rightarrow\bar{L}^\text{total}_{-n}$ if $S$ is an orthogonal matrix. Thus, the conformal condition (\ref{Virasorobndcond}) simplifies to
\begin{equation}\label{bosebndcon}
\paren{a^i_n-S_{ij}\,\bar{a}^j_{-n}}|\text{B}\rangle\rangle=0
\end{equation}
for $S$ an element of $O(2)$. This condition can also be constructed explicitly for free fields by requiring continuity of the stress tensor at the location of the interface \cite{Bachas:2001vj}. These new conditions (\ref{bosebndcon}) can be solved by a coherent state anzatz
\begin{equation}\label{coheranzatz}
|S\rangle\rangle=g\prod_{n=1}^\infty\exp\paren{\,\frac{1}{n}\,S_{ij}\,a^i_{-n}\bar{a}^j_{-n}}|\Omega\rangle
\end{equation}
The form of (\ref{bosebndcon}) describes a D-brane in the boundary state formalism (see \cite{DiVecchia:1999mal, DiVecchia:1999fje} for review), and this correspondence is used to find and classify all the possible boundary states for the two scalar model. The D-brane interpretation also gives us physical meaning for the normalization, the so called $g$-factor, and the ground state $|\Omega\rangle$ in (\ref{coheranzatz}).

\begin{figure}
\centering
\includegraphics[scale=1.5]{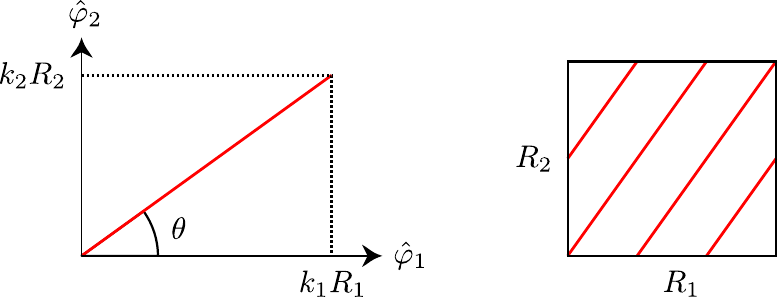}
\caption{On the right: A D1-brane wrapping the bosonic 2-torus continued into the compactification lattice so as to show the lattice intercept at $(k_1R_1,k_2R_2)$. On the left: A D1-brane wrapping the bosonic 2-torus (corresponding to the parameters $k_1=2$ and $k_2=3$) shown in the unit cell of the compactification lattice.\label{d1fig}}
\end{figure}

The one-dimensional special case of (\ref{bosebndcon}) emits the unit scalar choices $S=\pm 1$, which correspond to the two possible D-brane states for a single compact scalar
\begin{align}
|\text{D}0\rangle\rangle &= \sqrt{\frac{R}{\sqrt{2\alpha'}}}\,\prod_{n=1}^\infty\exp\paren{\frac{1}{n}\,a_{-n}\bar{a}_{-n}}\sum_{N=-\infty}^\infty e^{-iN\varphi_0/R}\,|N,0\rangle\label{boseD0}\\
|\text{D}1\rangle\rangle &= \sqrt{\frac{1}{R}\sqrt{\frac{\alpha'}{2}}}\,\prod_{n=1}^\infty\exp\paren{-\frac{1}{n}\,a_{-n}\bar{a}_{-n}}\sum_{M=-\infty}^\infty e^{iM\tilde{\varphi}_0}\,|0,M\rangle\label{boseD1}
\end{align}
respectively, where the D0-brane enforces a Dirichlet condition at the boundary and the D1-brane enforces a Neumann condition at the boundary. The constants $\varphi_0$ and $\tilde{\varphi}_0$ are position and dual Wilson line moduli of the D-brane. For an interface between two $c=1$ CFTs the D-brane states of the two scalar model are needed. These were constructed in \cite{Bachas:2007td} using rotations and T-duality transformations on the tensor products of (\ref{boseD0}) and (\ref{boseD1}). The first class of states are the rotations of
\begin{equation}
|\text{D}1,0\rangle\rangle=|\text{D}1\rangle\rangle\otimes|\text{D}0\rangle\rangle
\end{equation}
by an arbitrary angle in the compactification lattice parametrized by two integers $k_1$ and $k_2$
\begin{equation}\label{2boseangle}
\tan\theta=\frac{k_2R_2}{k_1R_1}
\end{equation}
as shown in figure \ref{d1fig}. The explicit boundary state is given by
\begin{equation}\label{D1theta}
|\text{D}1,\theta(k_1,k_2)\rangle\rangle = \sqrt{\frac{k_1^2R_1^2+k_2^2R_2^2}{2R_1R_2}}\,\prod_{n=1}^\infty\exp\paren{\frac{1}{n}\,S_{ij}(\theta)\,a^i_{-n}\bar{a}^j_{-n}}|\Omega\rangle
\end{equation}
where
\begin{equation}\label{d1S}
S(\theta)=\begin{pmatrix} \cos\theta & \sin\theta \\ -\sin\theta & \cos\theta \end{pmatrix}\begin{pmatrix} -1 & 0 \\ 0 & 1 \end{pmatrix}\begin{pmatrix} \cos\theta & -\sin\theta \\ \sin\theta & \cos\theta \end{pmatrix}=\begin{pmatrix} -\cos 2\theta & -\sin 2\theta \\ -\sin 2\theta & \cos 2\theta \end{pmatrix}
\end{equation}
and
\begin{equation}\label{D1thetazm}
|\Omega\rangle=\sum_{N,M=-\infty}^\infty e^{iN\alpha-iM\beta}|k_2N,k_1M\rangle\otimes |-k_1N,k_2M\rangle
\end{equation}
The other class of states, corresponding to bound states between $k_2$ D2-branes and $k_1$ D0-branes, is obtained from (\ref{D1theta}) through a T-duality transformation (\ref{Tdualrules}) of $\varphi_1$. Explicitly, the state is given by
\begin{equation}\label{2D2D0}
|k_2\text{D}2/k_1\text{D}0\rangle\rangle=\sqrt{\frac{k_1^2{\alpha'}^2+k_2^2R_1^2R_2^2}{2\alpha'R_1R_2}}\,\prod_{n=1}^\infty\exp\paren{\frac{1}{n}\,S'_{ij}(\theta')\,a^i_{-n}\bar{a}^j_{-n}}|\Omega'\rangle
\end{equation}
where
\begin{equation}
S'(\theta')=S(\theta')\begin{pmatrix} -1 & 0 \\ 0 & 1 \end{pmatrix}=\begin{pmatrix} \cos 2\theta' & -\sin 2\theta' \\ \sin 2\theta' & \cos 2\theta \end{pmatrix}
\end{equation}
with ``angle"
\begin{equation}\label{2boseangle2}
\tan\theta'=\frac{k_2R_1R_2}{k_1\alpha'}
\end{equation}
obtained from the replacement $R_1\rightarrow\alpha'/R_1$ in (\ref{2boseangle}), and
\begin{equation}
|\Omega'\rangle=\sum_{N,M=-\infty}^\infty e^{iN\alpha'-iM\beta'}|k_1M,k_2N\rangle\otimes |-k_1N,k_2M\rangle
\end{equation}
obtained from the replacement $n_1\leftrightarrow w_1$ in (\ref{D1thetazm}). The normalization factors in the previous boundary states are determined by Cardy's condition, which we will explain for a general bosonic D-brane state in the next section.

\subsection{Bosonic junctions}

\begin{figure}[t]
\centering
\includegraphics[scale=1.0]{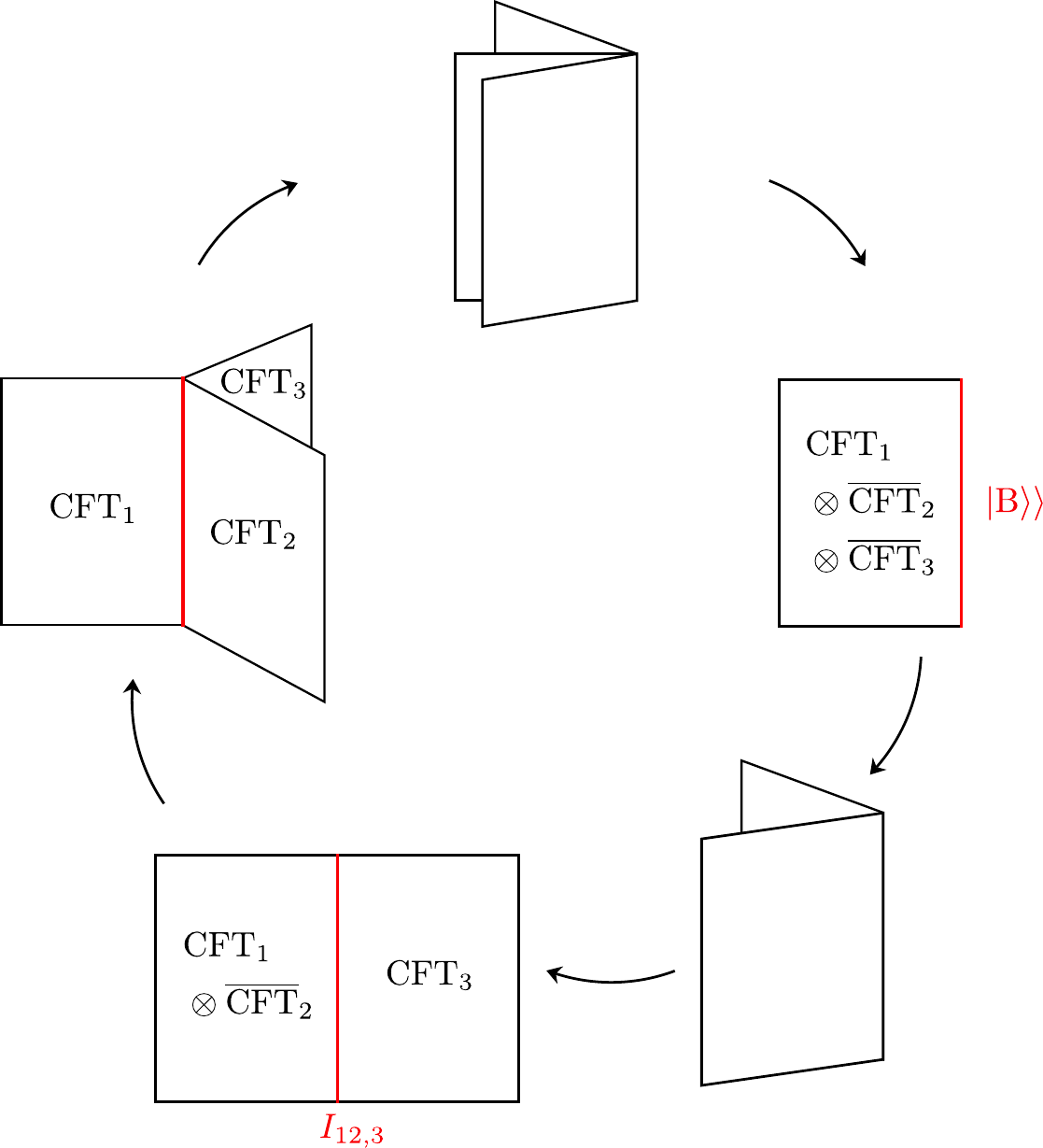}
\caption{Illustrating the unfolded, folded, and partially folded pictures for a 3-junction. As before, the folded picture is used to characterize the boundary states. However, for the entanglement entropy calculations we will only unfold one CFT and work with interface operators in this partially folded picture.}\label{juncfoldfig}
\end{figure}

For junctions connecting $N>2$ free boson CFTs, we proceed with the same folding methods shown in figure \ref{intfoldfig} applied repeatedly, as illustrated in figure \ref{juncfoldfig}. Specifically, the bosonic $N$-junction is folded into the $N$-times tensor product CFT with boundary states $|\text{B}\rangle\rangle$ determined by the boundary condition 
\begin{equation}\label{Nbndcond}
\paren{a^i_n-S_{ij}\,\bar{a}^j_{-n}}|\text{B}\rangle\rangle=0
\end{equation}
where now $S$ is an element of $O(N)$\footnote{This is seen either by the easily generalized replacement in (\ref{opreplace}) or by requiring continuity of the stress tensor at the location of the junction \cite{Chiodaroli:2010mv}.}. As before, (\ref{Nbndcond}) is solved by a coherent state of the form
\begin{equation}\label{NbosBstate}
|S\rangle\rangle=g\prod_{n=1}^\infty\exp\paren{\frac{1}{n}\,S_{ij}\,a^i_{-n}\bar{a}^j_{-n}}\sum_{(\mathbf{a}_0,\bar{\mathbf{a}}_0)\in\Lambda}e^{i\delta_{\mathbf{a}_0,\bar{\mathbf{a}}_0}}\bigotimes_{i=1}^N|n_i,w_i\rangle
\end{equation}
where $\Lambda$ is an $N$-dimensional sublattice of the full $2N$-dimensional lattice of unconstrained eigenvalues of the $a^i_0$ and $\bar{a}^i_0$. Not every element of $O(N)$ will be compatible with the zero mode structure, i.e. satisfy the $n=0$ case of (\ref{Nbndcond}) for the quantized eigenvalues (\ref{zmeigen}), and thus the bosonic boundary states correspond to a countable subset of $O(N)$. For $N=2$ the restrictions (\ref{2boseangle}) and (\ref{2boseangle2}) specify the allowed subset of $O(2)$, and in section \ref{sec5} we find the allowed subset of $O(3)$ for $N=3$. Lastly, the phases $\delta_{\mathbf{a}_0,\bar{\mathbf{a}}_0}$ are related to the position and dual Wilson line moduli of the D-brane, but as they will vanish from all our calculations we will not characterize them further.

We now fix the normalization through Cardy's condition for this general bosonic D-brane. Cardy's condition enforces the consistency between the open and closed string channels; that is, it requires the annulus amplitude to have a modular interpretation as a partition function on the cylinder. We will use this condition to fix the value of the normalization factor in (\ref{NbosBstate}). Let $q=e^{-2\pi t}$ for some $t>0$. The annulus amplitude is then
\begin{equation}\label{annamp}
\langle\langle S|q^{\,\sum_{i=1}^N(L^i_0+\bar{L}^i_0-1/12)}|S\rangle\rangle
\end{equation}
The quadratic operator exponentials in the boundary state complicate attempts at direct calculation; instead we linearize the exponential by means of Gaussian integrals of the form
\begin{equation}\label{boslinint}
e^{\mathbf{A}\cdot\mathbf{B}}=\int\frac{d^N\mathbf{z}\,d^N\bar{\mathbf{z}}}{\pi^N}\,\,e^{-\mathbf{z}\cdot\bar{\mathbf{z}}-\mathbf{z}\cdot\mathbf{A}-\bar{\mathbf{z}}\cdot\mathbf{B}}
\end{equation}
where $\mathbf{A}$ and $\mathbf{B}$ are $N$-dimensional vectors whose entries are all mutually commuting operators. Linearizing each of the exponentials in (\ref{annamp}) with (\ref{boslinint}) in a complementary fashion we obtain the expression
\begin{align}\label{annamplin}
\langle\langle S|q^{\,\sum_{i=1}^N(L^i_0+\bar{L}^i_0-1/12)}|S\rangle\rangle &= g^2\,\langle\Omega|q^{\,\sum_{i=1}^N(L^i_0+\bar{L}^i_0)}|\Omega\rangle\nonumber\\
&\times q^{-N/12}\prod_{n,m=1}^\infty\int\frac{d^N\mathbf{z}_nd^N\bar{\mathbf{z}}_nd^N\mathbf{w}_md^N\bar{\mathbf{w}}_m}{\pi^{2N}}\,\,e^{-\mathbf{z}_n\cdot\bar{\mathbf{z}}_n-\mathbf{w}_m\cdot\bar{\mathbf{w}}_m}\\
&\times\langle 0|e^{-q^m\mathbf{w}_m\cdot S^\text{T}\mathbf{a}_m-\tfrac{1}{m}q^m\bar{\mathbf{w}}_m\cdot\bar{\mathbf{a}}_m}e^{-\tfrac{1}{n}\mathbf{z}_n\cdot\mathbf{a}_{-n}-\bar{\mathbf{z}}_n\cdot S\bar{\mathbf{a}}_{-n}}|0\rangle\nonumber
\end{align}
where $|\Omega\rangle$ is the lattice-summed zero mode in (\ref{NbosBstate}) and we have used the identities
\begin{equation}\label{boseViraident}
e^{a_n}q^{L_0} = q^{L_0}e^{q^na_n}\,\,\,\,\,\,\text{and}\,\,\,\,\,\,e^{\bar{a}_n}q^{\bar{L}_0} = q^{\bar{L}_0}e^{q^n\bar{a}_n}
\end{equation}
The form of (\ref{annamplin}) is such that the zero mode contribution, the first line of (\ref{annamplin}), is isolated from the remaining oscillator contribution. The zero mode contribution is a lattice theta function (see appendix \ref{thetaS})
\begin{equation}
g^2\,\langle\Omega|q^{\,\sum_{i=1}^N(L^i_0+\bar{L}^i_0)}|\Omega\rangle = g^2\,\Theta_\Lambda(2it)
\end{equation}
where the dependence on the phases in $|\Omega\rangle$ have vanished. For the oscillator integrals, we commute the two linear operator exponentials in the third line of (\ref{annamplin}) to obtain
\begin{align}
&q^{-N/12}\prod_{n=1}^\infty\int\frac{d^N\mathbf{z}_nd^N\bar{\mathbf{z}}_nd^N\mathbf{w}_nd^N\bar{\mathbf{w}}_n}{\pi^{2N}}\,\,e^{-\mathbf{z}_n\cdot\bar{\mathbf{z}}_n-\mathbf{w}_n\cdot\bar{\mathbf{w}}_n+q^n\mathbf{z}_n\cdot S\mathbf{w}_n+q^n\bar{\mathbf{z}}_n\cdot S\bar{\mathbf{w}}_n}\\
&= q^{-N/12}\prod_{n=1}^\infty\int\frac{d^N\mathbf{w}_nd^N\bar{\mathbf{w}}_n}{\pi^{N}}\,\,e^{-(1-q^{2n})\,\mathbf{w}_n\cdot\bar{\mathbf{w}}_n} = \Big[q^{1/12}\prod_{n=1}^\infty\paren{1-q^{2n}}\Big]^{-N}
\end{align}
where the dependence on $S$ is removed after the $\mathbf{z}_n$, $\bar{\mathbf{z}}_n$ integration due to the fact that $S^\text{T}S=1_N$ as $S$ is an element of $O(N)$. Comparing this result to (\ref{Dedetadef}) we find that the annulus amplitude can be written in closed form as
\begin{equation}
\langle\langle S|q^{\,\sum_{i=1}^N(L^i_0+\bar{L}^i_0-1/12)}|S\rangle\rangle = g^2\,\Theta_\Lambda(2it)\,[\eta(2it)]^{-N}
\end{equation}
Performing $S$-transformations on the above we have the equivalent expression
\begin{equation}\label{Ncylpar}
\langle\langle S|q^{\,\sum_{i=1}^N(L^i_0+\bar{L}^i_0-1/12)}|S\rangle\rangle = \frac{g^2}{\text{vol}(\Lambda)}\,\Theta_{\Lambda^*}(i/2t)\,[\eta(i/2t)]^{-N}
\end{equation}
In order for (\ref{Ncylpar}) to correspond to a cylinder partition function with a properly normalized vacuum we must have that the constant term as $t\rightarrow 0$ in (\ref{Ncylpar}) is unity. Thus, Cardy's condition fixes
\begin{equation}
g=\sqrt{\text{vol}(\Lambda)}
\end{equation}

\subsection{Fermionic interfaces and junctions}

Owing to their much less complicated zero mode structure, the boundary states corresponding to interfaces and junctions between free fermion CFTs have a simpler construction and can be expressed entirely in terms of an arbitrary element of $O(N)$. The fermionic analog to (\ref{bosebndcon}) is
\begin{equation}
\paren{\psi^i_n+iS_{ij}\bar{\psi}^j_{-n}}|\text{B}\rangle\rangle=0
\end{equation}
In contrast to (\ref{boseD0}) and (\ref{boseD1}) the single fermion has the four possible boundary states
\begin{align}
|\epsilon\rangle\rangle_\text{NS} &= \prod_{n\in\mathbb{N}-\tfrac{1}{2}}\exp\paren{i\epsilon\psi_{-n}^i\bar{\psi}^j_{-n}}|0\rangle\\
|\epsilon\rangle\rangle_\text{R} &= 2^{\tfrac{1}{4}}\prod_{n=1}^\infty\exp\paren{i\epsilon\psi_{-n}^i\bar{\psi}^j_{-n}}|\epsilon\rangle
\end{align}
corresponding to $\epsilon=\pm 1$ and the different modings in the Neveu-Schwarz and Ramond sectors. Each of these boundary states are normalized via Cardy's condition as in the bosonic case. In \cite{Bachas:2012bj} the various fermionic boundary states for $N=2$ were found; here we give their straightforward generalization to arbitrary $N$ for the Neveu-Schwarz sector
\begin{equation}\label{NfermBstate}
|S\rangle\rangle_\text{NS}=\prod_{n\in\mathbb{N}-\tfrac{1}{2}}\exp\paren{iS_{ij}\psi_{-n}^i\bar{\psi}^j_{-n}}\bigotimes_{i=1}^N|0\rangle
\end{equation}
which will be the focus of the fermionic calculations in this work, and for the Ramond sector
\begin{equation}\label{NfermRBstate}
|S\rangle\rangle_\text{R}=\sqrt{\frac{2}{\det\paren{1-\mathcal{F}}}}\,\prod_{n=1}^\infty\exp\paren{iS_{ij}\psi_{-n}^i\bar{\psi}^j_{-n}}\exp\paren{\frac{1}{2}\,\mathcal{F}_{ij}\gamma_{-\epsilon_i}^i\gamma_{-\epsilon_j}^j}\bigotimes_{i=1}^N|\epsilon_i\rangle
\end{equation}
where
\begin{equation}
\gamma^i_\pm=\frac{1}{\sqrt{2}}\paren{\psi^i_0\pm i\bar{\psi}^i_0}
\end{equation}
and $\mathcal{F}$ is an anti-symmetric matrix given by
\begin{equation}
S'=\paren{1_N+\mathcal{F}}^{-1}\paren{1_N-\mathcal{F}}\,\,\Longleftrightarrow\,\,\mathcal{F}=\paren{1_N-S'}^{-1}\paren{1_N+S'}
\end{equation}
The state in (\ref{NfermRBstate}) is only well defined as long as $S'$ is in the connected component of $O(N)$. Thus we take the matrix $S'$ to be the pure rotation part of $S$, i.e. we write $S$ as an elementary reflection composed with a continuous rotation $S'$. The reflection content of $S$ is then represented in the ground state through the choice of signs in the $\epsilon_i$. If $S$ is a pure rotation then $\epsilon_i=+1$ for all $i$, whereas if $S$ includes a reflection then $\epsilon_i=-1$ for all $i$ excepting the two indices corresponding to the plane of reflection. These considerations ensure that (\ref{NfermRBstate}) satisfies the zero mode boundary condition
\begin{equation}
\paren{\psi_0^i+iS_{ij}\bar{\psi}_0^j}|S\rangle\rangle_\text{R}=0\,\,\Longleftrightarrow\,\,\paren{\gamma^i_{\epsilon_i}+\mathcal{F}_{ij}\gamma^j_{-\epsilon_j}}|S\rangle\rangle_\text{R}=0
\end{equation}
while maintaining a finite normalization.

\subsection{Reflection and transmission}

In \cite{Sakai:2008tt} and \cite{Brehm:2015lja} it was shown that the physical quantity determining the universal term in the entanglement entropy for both the bosonic and fermionic interfaces is the transmission coefficient of the interface. This continues to be the case for $N>2$, so therefore we briefly review these coefficients for interfaces and junctions of free boson and free fermion CFTs.

The reflection and transmission coefficients for CFT $N$-junctions are related to the $N\times N$ matrix
\begin{equation}
R_{ij}=\frac{\langle 0| L_2^i\bar{L}_2^j|\text{B}\rangle\rangle}{\langle 0|\text{B}\rangle\rangle}
\end{equation}
where $|\text{B}\rangle\rangle$ is the boundary state corresponding to the junction. This matrix was first considered for interfaces in \cite{Quella:2006de} where average reflection and transmission coefficients were found
\begin{equation}\label{reftrans2}
\mathcal{R}^\text{avg}=\frac{2}{c_1+c_2}\paren{R_{11}+R_{22}}\,\,\,\,\,\,\text{and}\,\,\,\,\,\,\mathcal{T}^\text{avg}=\frac{2}{c_1+c_2}\paren{R_{12}+R_{21}}
\end{equation}
which are enough to characterize transport processes for $N=2$ since in this case $R$ is a symmetric matrix. These coefficients were generalized in \cite{Kimura:2015nka} to the case $N\geq 2$
\begin{equation}\label{reftransN}
\mathcal{R}_i=\frac{2}{c_i}\,R_{ii}\,\,\,\,\,\,\text{and}\,\,\,\,\,\,\mathcal{T}_{ij}=\frac{2}{c_i}\,R_{ij}
\end{equation}
where $\mathcal{R}_i$ is the reflection coefficient for CFT$_i$ and $\mathcal{T}_{ij}$ is the transmission coefficient for transport from CFT$_i$ to CFT$_j$. It should be noted for $N=2$ that (\ref{reftransN}) is related to (\ref{reftrans2}) by
\begin{equation}
\mathcal{T}_{12}=\frac{c_2}{c_1}\,\mathcal{T}_{21}=\frac{c_1+c_2}{2c_1}\,\mathcal{T}^\text{avg}
\end{equation}
so that for $c_1=c_2=c$ the three different transmissions all agree. For $N>2$ we'll also want to consider the total transmission from CFT$_i$, given by the sum
\begin{equation}
\mathcal{T}_i=\sum_{j\neq i}\mathcal{T}_{ij}
\end{equation}
In both the free boson and free fermion cases (\ref{NbosBstate}) and (\ref{NfermBstate}), the reflection and transmission coefficients of these boundary states are given by
\begin{equation}\label{junctransref}
\mathcal{R}_i=S_{ii}^2\,\,\,\,\,\,\text{and}\,\,\,\,\,\,\mathcal{T}_{ij}=S_{ij}^2\,\,\Longrightarrow\,\,\mathcal{T}_i=1-S_{ii}^2
\end{equation}
and thus the coefficients can be lifted from the matrix $S$, e.g. the angled D1-brane with matrix (\ref{d1S}) has a transmission coefficient
\begin{equation}
\mathcal{T}=\sin^22\theta
\end{equation}

It is interesting to note that a completely transmissive junction, which necessarily has $\mathcal{R}_i=0$ for all $i=1,\ldots,N$, has its transmission coefficients constrained to be
\begin{equation}
\mathcal{T}_{ij}=\delta_{jk_i}
\end{equation}
where $k_{i+1}=k_i+1$, the index $N+1$ is identified with 1, and $k_{i}\neq i$. These correspond to twisted permutation junctions whose boundary states satisfy
\begin{equation}
a^i_{n}|S\rangle\rangle = \pm \bar{a}^{k_i}_{-n}|S\rangle\rangle
\end{equation}
for (\ref{NbosBstate}) and
\begin{equation}
\psi^i_n|S\rangle\rangle = \pm i\bar{\psi}^{k_i}_{-n}|S\rangle\rangle
\end{equation}
for (\ref{NfermBstate}) with independent sign choices for each $i$, of which there are $2^N(N-1)$ distinct matrices $S$.

\section{Entanglement entropy at conformal interfaces}
\label{sec3}

Here we review the entanglement entropy calculations of \cite{Sakai:2008tt} and \cite{Brehm:2015lja} for interfaces between free boson and free fermion CFTs. We choose to first highlight the bosonic calculation as it will be the one most readily generalizable to arbitrary $N$. In section \ref{sec2} the starting point for characterizing an interface was to consider the corresponding boundary state in the folded picture. Once the boundary state is obtained the folded CFT must then be unfolded to produce the interface operator satisfying (\ref{intcond}) that is needed for the calculation.

The bosonic boundary states in (\ref{D1theta}) and (\ref{2D2D0}) are unfolded into operators via what is essentially a parity transformation on the quantities of one of the CFTs \cite{Bachas:2007td}
\begin{equation}\label{boseunfold}
|n,w\rangle\longrightarrow \langle -n,w|\,,\,\,\,\,\,\,a_{-n}\longrightarrow -\bar{a}_{n}\,,\,\,\,\,\,\,\bar{a}_{-n}\longrightarrow -a_{n}
\end{equation}
Choosing to unfold $\varphi_2$ for the state (\ref{D1theta}) produces the interface operator
\begin{equation}\label{boseintop}
I_{1,2}=G_{1,2}\prod_{n=1}^\infty\exp\left\{\,\frac{1}{n}\big[S_{11}(\theta)\,a_{-n}^1\bar{a}^1_{-n}-S_{12}(\theta)\,a^1_{-n}a^2_n-S_{21}(\theta)\,\bar{a}^2_n\bar{a}^1_{-n}+S_{22}(\theta)\,\bar{a}^2_na^2_n\,\big]\right\}
\end{equation}
where the ground state operator given by
\begin{equation}
G_{1,2}=\sqrt{\frac{k_1^2R_1^2+k_2^2R_2^2}{2R_1R_2}}\sum_{N,M=-\infty}^\infty e^{iN\alpha-iM\beta}|k_2N,k_1M\rangle\langle k_1N,k_2M|
\end{equation}
The expression for the interface operator in (\ref{boseintop}) is a formal one, as the negatively-moded oscillators must be placed on the left side of the ground state operator after the full expansion of the exponential. An explicit expression for the interface operator can be obtained by a linearization of the exponential as in (\ref{boslinint}), one such choice being
\begin{align}\label{bose2intop}
I_{1,2} &= \prod_{n=1}^\infty\int\frac{d^2\mathbf{z}_n\,d^2\bar{\mathbf{z}}_n}{\pi^2}\,\,e^{-\mathbf{z}_n\cdot\bar{\mathbf{z}}_n}e^{-\frac{1}{n}z_{n1}a^1_{-n}-(S_{11}\bar{z}_{n1}-S_{21}\bar{z}_{n2})\,\bar{a}^1_{-n}}\nonumber\\
&\,\,\,\,\,\,\,\,\,\,\,\,\,\,\,\,\,\,\,\,\,\times G_{1,2}\prod_{n=1}^\infty e^{-\frac{1}{n}z_{n2}\bar{a}^2_{n}-(S_{22}\bar{z}_{n2}-S_{12}\bar{z}_{n1})\,a^2_{n}}
\end{align}

\begin{figure}
\centering
\includegraphics[scale=1.5]{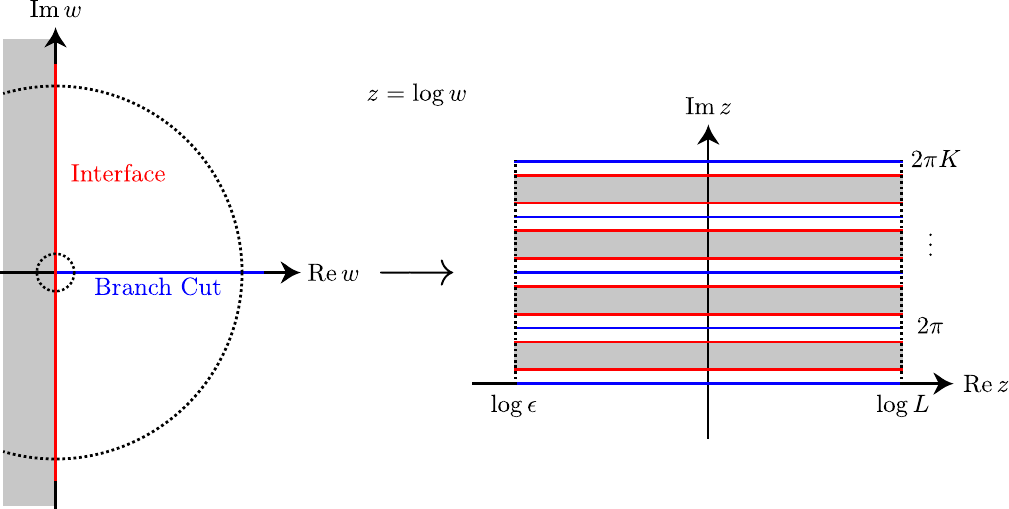}
\caption{The logarithmic map $z=\log w$ maps the $K$-sheeted Riemann surface, a single branch of which is shown on the left, to the geometry on the right. The circles on the left part of the figure correspond to an UV cutoff located $|w|=\epsilon$ and an IR cutoff located at $|w|=L$, with their image under the mapping forming the negative and positive real boundaries of the geometry on the right. This figure was adapted from \cite{Brehm:2015lja}.\label{replicafig}}
\end{figure}

With expressions for the interface operator like the above the entanglement entropy can be calculated through a geometric replica trick first formulated in \cite{Holzhey:1994we}, which is illustrated in figure \ref{replicafig}. The entanglement entropy is calculated as a limit of Renyi entropies of the reduced density matrix
\begin{equation}\label{SEEfromRenyi}
\mathcal{S}=-\frac{\partial}{\partial K}\,\text{Tr}_1\rho_1^K\Big|_{K=1}
\end{equation}
The trace of the $K$-th power of the reduced density matrix is re-written as a partition function on a $K$-sheeted Riemann surface $\mathcal{R}_K$ whose branch cut runs along a time-slice of CFT$_1$. From the path integral form
\begin{equation}
Z(K)=\int\mathcal{D}\varphi_1\mathcal{D}\varphi_2\,\exp\left[-\int_{\mathcal{R}_K}dw\,d\bar{w}\,\mathcal{L}(\varphi_1,\varphi_2)\right]
\end{equation}
the entanglement entropy in (\ref{SEEfromRenyi}) can be written in terms of this replicated partition function
\begin{equation}\label{SEEfromZK}
\mathcal{S}=\paren{1-\partial_K}\log Z(K)\,\big|_{K=1}
\end{equation}
Cutting off the $w$-plane outside the annulus $\epsilon<|w|<L$, the mapping $z=\log w$ maps this $K$-sheeted region into a rectangular region in the $z$-plane with $\text{Im}\,z=0$ and $\text{Im}\,z=2\pi K$ identified. For ease of calculation we further identify $\text{Re}\,z=\log\epsilon$ and $\text{Re}\,z=\log L$ so that the replicated partition function becomes the torus partition function with $2K$ interfaces inserted
\begin{equation}\label{Zkdef}
Z(K)=\text{Tr}_1\left[\big(I_{1,2}\,q^{H_2}I_{1,2}^\dagger\,q^{H_1}\big)^K\right]
\end{equation}
for $q=e^{-2\pi t}$ with $t=\pi/\log(L/\epsilon)$ after a rescaling of the $z$-plane (see \cite{Sakai:2008tt} for details). Combined with explicit interface operator expressions like (\ref{bose2intop}), the operator expression in (\ref{Zkdef}) can be used to calculate the exact form of the replicated partition function.

Calculating the commutation of the various operators between the ground state operators of successive interfaces, the partition function (\ref{Zkdef}) is written as a $2K$-(complex) dimensional  Gaussian integral. Thus the final evaluation of $Z(K)$ is performed through calculation of a determinant and re-expressed in terms of modular functions
\begin{align}
Z(K) &= g^{2K}K|\sin 2\theta|^{K-1}\theta_3\paren{\frac{itKk_2^2\alpha'}{R_1^2\sin^2\theta}}\theta_3\paren{\frac{itKk_1^2R_1^2}{\alpha'\cos^2\theta}}[\eta(2it)]^{K-3}\prod_{k=1}^{K-1}\theta_1^{-1}(\nu_k|2it)\label{prevZK}\\
&= \text{vol}(\Lambda)^KK\,\mathcal{T}^{\,(K-1)/2}\,\Theta_\Lambda(2iKt)\left[\eta\paren{2it}\right]^{K-3}\prod_{k=1}^{K-1}\theta^{-1}_1(\nu_k|2it)\label{newZK}
\end{align}
where
\begin{equation}
\sin\pi\nu_k=|\sin 2\theta|\sin\frac{\pi k}{K}=\sqrt{\mathcal{T}}\,\sin\frac{\pi k}{K}
\end{equation}
The form of the partition function in (\ref{prevZK}) is the one given in \cite{Sakai:2008tt}, whereas the form in (\ref{newZK}) uses conventions more readily comparable to the $N>2$ cases. The remaining product in the partition function is analytically continued in $K$, which is reviewed in appendix \ref{bernpolys}, so that from (\ref{SEEfromZK}) the entanglement entropy is
\begin{equation}\label{bose2EE}
\mathcal{S}=\frac{1}{2}\,\sigma\big(|\sin 2\theta|\big)\log\frac{L}{\epsilon}-\log|k_1k_2|
\end{equation}
with the function $\sigma(s)$ in (\ref{sigmadef}). The function $\sigma(s)$ increases monotonically from $\sigma(0)=0$ to $\sigma(1)=1/3$, matching the behavior of the universal term expected of the entanglement entropy of a semi-infinite interval in a $c=1$ CFT as discussed in section \ref{sec1}.

The entanglement entropy of the fermionic interface follows the same general procedure as the bosonic interface calculation, i.e. inserting the unfolded interface operators into (\ref{Zkdef}) in order to calculate (\ref{SEEfromZK}). The $N=2$ fermionic boundary states of (\ref{NfermBstate}) and (\ref{NfermRBstate}) are unfolded into operators via the transformation \cite{Bachas:2007td}
\begin{equation}\label{fermunfold}
|0\rangle\,\,\longrightarrow\,\,\langle 0|\,,\,\,\,\,\,\,|\epsilon\rangle\,\,\longrightarrow\,\,\langle\epsilon|\,,\,\,\,\,\,\,\psi_n\,\,\longrightarrow\,\,-i\bar{\psi}_{-n}\,,\,\,\,\,\,\,\bar{\psi}_n\,\,\longrightarrow\,\,i\psi_{-n}
\end{equation}
For the fermionic interfaces the explicit expansion of the quadratic operator exponential is considerably simpler than in the bosonic interfaces due to the fact that for each fixed mode $n$ the Hilbert space $\mathcal{H}_n$ of the corresponding fermionic oscillator is 4-dimensional (as opposed to the infinite-dimensional situation for the bosonic oscillators). As such, the matrix representation on the ordered basis $\{\psi_{-n}|0\rangle,\bar{\psi}_{-n}|0\rangle,\psi_{-n}\bar{\psi}_{-n}|0\rangle,|0\rangle\}$ is
\begin{equation}
I_{1,2}=\left\{\prod_{n>0}I^n_{1,2}\right\}I^0_{1,2}
\end{equation}
where
\begin{equation}
I^n_{1,2}=\begin{pmatrix} S_{12} & 0 & 0 & 0 \\ 0 & S_{21} & 0 & 0 \\ 0 & 0 & -\det S & -iS_{11} \\ 0 & 0 & -iS_{22} & 1 \end{pmatrix}
\end{equation}
The partition function is then calculated in terms of the four eigenvalues $\lambda_{j,n}$ of the block matrix
\begin{equation}
I^{n}_{1,2}P^n_2\paren{I^n_{1,2}}^\dagger P^n_1
\end{equation}
where matrix representations of the propagators are
\begin{equation}
P^n_i=\begin{pmatrix} q^n & 0 & 0 & 0 \\ 0 & q^n & 0 & 0 \\ 0 & 0 & q^{2n} & 0 \\ 0 & 0 & 0 & 1 \end{pmatrix}
\end{equation}
Explicitly for the NS interface, the partition function in terms of the eigenvalues can be re-expressed in terms of modular functions
\begin{align}
Z(K) &= \prod_{n\in\mathbb{N}-\tfrac{1}{2}}\paren{\lambda_{1,n}^K+\lambda_{2,n}^K+\lambda_{3,n}^K+\lambda_{4,n}^K} = \frac{\theta_3(2it)}{\left[\eta(2it)\right]^K}\prod_{k=1}^{K-1}\theta_3(\nu_k|2it) \label{fermZKmod}
\end{align}
by utilizing the algebraic identity\footnote{From the form of (\ref{mainalgident}) it appears that the final equality in (\ref{fermZKmod}) is only valid for odd values of $K$. In \cite{Brehm:2015lja} it was shown that this suffices for calculating the entanglement entropy. Interestingly enough, we will later show that the expression is valid for even $K$ as well.}
\begin{equation}\label{mainalgident}
\prod_{k=1}^{K-1}\left[x^2-2xy\cos\paren{\theta+\frac{2\pi k}{K}}+y^2\right]=x^{2K}-2x^Ky^K\cos\paren{K\theta}+y^{2K}
\end{equation}
The analytic continuation in $K$ is similar to the bosonic case, and the entanglement entropy is
\begin{equation}\label{ferm2EE}
\mathcal{S}=\frac{1}{2}\,\left[\frac{1}{2}\sqrt{1-S_{11}^2}-\sigma\big(\sqrt{1-S_{11}^2}\,\big)\right]\log\frac{L}{\epsilon}
\end{equation}
with the universal term satisfying the same limiting behavior as (\ref{bose2EE}) for a $c=1/2$ CFT.

\section{Entanglement entropy at $N$-junctions}
\label{sec4}

The starting point for the junction entanglement calculations is the same as in the interface case: with the corresponding boundary state $|\text{B}\rangle\rangle$ in the folded picture (see figures \ref{intfoldfig} and \ref{juncfoldfig}). For the interfaces the tensor product CFT is then unfolded to obtain the interface operator $I_{1,2}$ to be used in calculating the replicated partition function (\ref{Zkdef}). This same basic strategy can be applied to the junction case as well by noting that it is equivalent to replacing in CFT$_1$ with $\bigotimes_{j\neq i}\text{CFT}_j$ and CFT$_2$ with CFT$_i$ in figure \ref{intfoldfig}. This is the partially folded picture (shown in figure \ref{juncfoldfig} for $N=3$) where, for the purposes of calculating the entanglement entropy of CFT$_i$, we only need an interface operator $I_{1\ldots N,i}$ taking states from CFT$_i$ to the rest of the CFTs in the junction as a tensor product. Thus, the replicated partition function has essentially the same from as (\ref{Zkdef}); that is
\begin{equation}\label{ZKNdef}
Z(K)=\text{Tr}_{1\ldots N}\left[\paren{I_{1\ldots N,i}\,q^{H_i}(I_{1\ldots N,i})^\dagger q^{H_{1\ldots N}}}^K\right]
\end{equation}
where $H_{1\ldots N}$ is the Hamiltonian of $\bigotimes_{j\neq i}\text{CFT}_j$.

\subsection{Bosonic junction}

We'll begin our calculations with the bosonic boundary state (\ref{NbosBstate}). Unfolding the $i$-th boson according to (\ref{boseunfold}), we linearize via (\ref{boslinint}) in order to obtain explicit expressions for the interface and anti-interface operators
\begin{align}
I_{1\ldots N,i} &= \prod_{n=1}^\infty\int\frac{d^N\mathbf{z}_n\,d^N\bar{\mathbf{z}}_n}{\pi^N}\,\,e^{-\mathbf{z}_n\cdot\bar{\mathbf{z}}_n-\frac{1}{n}\sum_{j\neq i}z_{nj}a^j_{-n}-\sum_{j\neq i}\sum_lS_{lj}\bar{z}_{nl}\bar{a}^j_{-n}}\nonumber\\
&\,\,\,\,\,\,\,\,\,\,\,\,\,\,\,\,\,\,\,\,\,\times G_{1\ldots N,i}\prod_{n=1}^\infty e^{\,\frac{1}{n}z_{ni}\bar{a}^i_n+\sum_lS_{li}\bar{z}_{nl}a^i_n}\label{boseNintop}\\
(I_{1\ldots N,i})^\dagger &= \prod_{n=1}^\infty\int\frac{d^N\mathbf{w}_n\,d^N\bar{\mathbf{w}}_n}{\pi^N}\,\,e^{-\mathbf{w}_n\cdot\bar{\mathbf{w}}_n+\sum_{l}S_{il}w_{nl}\bar{a}^i_{-n}+\frac{1}{n}\bar{w}_{ni}a^i_{-n}}\nonumber\\
&\,\,\,\,\,\,\,\,\,\,\,\,\,\,\,\,\,\,\,\,\,\times (G_{1\ldots N,i})^\dagger\prod_{n=1}^\infty e^{-\sum_{j\neq i}\sum_lS_{jl}w_{nl}a^j_n-\frac{1}{n}\sum_{j\neq i}\bar{w}_{nj}\bar{a}^j_n}\label{boseNaintop}
\end{align}
with the ground state operator given by
\begin{equation}\label{boseGopdef}
G_{1\ldots N,i}=\sqrt{\text{vol}(\Lambda)}\sum_{(\mathbf{a}_0,\bar{\mathbf{a}}_0)\in\Lambda}e^{i\delta_{\mathbf{a}_0,\bar{\mathbf{a}}_0}}\Big(\bigotimes_{j\neq i}|n_j,w_j\rangle\Big)\otimes\langle -n_i,w_i|
\end{equation}
which are needed to compute the partition function (\ref{ZKNdef}). From (\ref{boseNintop}) and (\ref{boseNaintop}) we then calculate the commutation between the various exponentials of the oscillators of the $i$-th boson in the relevant partition function block
\begin{align}
J &= q^{-N/12}I_{1\ldots N,i}\,q^{L^i_0+\bar{L}^i_0}\paren{I_{1\ldots N,i}}^\dagger q^{\,\sum_{j\neq i} (L^j_0+\bar{L}^j_0)}\label{boseJdef}\\
&= \prod_{n=1}^\infty\int\frac{d^N\mathbf{z}_n\,d^N\bar{\mathbf{z}}_n\,d^N\mathbf{w}_n\,d^N\bar{\mathbf{w}}_n}{\pi^{2N}}\,\,e^{-\mathbf{z}_n\cdot\bar{\mathbf{z}}_n -\mathbf{w}_n\cdot\bar{\mathbf{w}}_n+q^n\sum_l\paren{S_{il}z_{ni}w_{nl}+S_{li}\bar{z}_{nl}\bar{w}_{ni}}}\,\mathcal{O}_LG'\mathcal{O}_R\label{boseJexp}
\end{align}
where the remaining oscillators are contained in
\begin{align}
\mathcal{O}_L &= \prod_{n=1}^\infty\exp\Big[-\frac{1}{n}\sum_{j\neq i}z_{nj}a^j_{-n}-\sum_{j\neq i}\sum_lS_{lj}\bar{z}_{nl}\bar{a}^j_{-n}\Big]\label{boseOLdef}\\
\mathcal{O}_R &= \prod_{n=1}^\infty\exp\Big[-q^n\Big(\sum_{j\neq i}\sum_lS_{jl}w_{nl}a^j_n+\frac{1}{n}\sum_{j\neq i}\bar{w}_{nj}\bar{a}^j_{n}\Big)\Big]\label{boseORdef}
\end{align}
and the zero mode information is encoded in the operator
\begin{equation}\label{boseGpdef}
G'=\text{vol}(\Lambda)\,q^{-N/12}\sum_{(\mathbf{a}_0,\bar{\mathbf{a}}_0)\in\Lambda}q^{|\mathbf{a}_0|^2+|\bar{\mathbf{a}}_0|^2}\Big(\bigotimes_{j\neq i}|n_j,w_j\rangle\Big)\otimes\Big(\bigotimes_{j\neq i}\langle n_j,w_j|\Big)
\end{equation}
Notice that in the above that the phases $\delta_{\mathbf{a}_0,\bar{\mathbf{a}}_0}$ originally present in (\ref{boseGopdef}) have vanished from the calculation. Also, the additional factors of $q^n$ in (\ref{boseORdef}) and the weighting of the lattice sum in (\ref{boseGpdef}) result from the identity (\ref{boseViraident}) and the application of the propagators on the vacuum states in (\ref{boseGopdef}).

Using the expression (\ref{boseJexp}) for the block (\ref{boseJdef}), we can now write the $K$-sheeted partition function (\ref{ZKNdef}) in terms of this block
\begin{align}
Z(K) &= \text{Tr}_{1\ldots N}\paren{J^K}\\
&= \prod_{n=1}^\infty\int\prod_{k=1}^K \frac{d^N\mathbf{z}_n^{(k)}d^N\bar{\mathbf{z}}_n^{(k)}d^N\mathbf{w}_n^{(k)}d^N\bar{\mathbf{w}}_n^{(k)}}{\pi^{2N}}\,\,e^{-\mathbf{z}^{(k)}_n\cdot\bar{\mathbf{z}}^{(k)}_n -\mathbf{w}^{(k)}_n\cdot\bar{\mathbf{w}}^{(k)}_n+q^n\sum_l\paren{S_{il}z^{(k)}_{ni}w^{(k)}_{nl}+S_{li}\bar{z}^{(k)}_{nl}\bar{w}^{(k)}_{ni}}}\nonumber\\
&\,\,\,\,\,\,\,\,\,\,\,\,\,\,\,\,\,\,\,\,\times\text{Tr}_{1\ldots N}\paren{G'\mathcal{O}^{(1)}_R\mathcal{O}^{(2)}_L G'\mathcal{O}^{(2)}_R\cdots\mathcal{O}^{(K)}_LG'\mathcal{O}^{(K)}_R\mathcal{O}^{(1)}_L}\label{boseZtrace}\\
&=\text{vol}(\Lambda)^{K}q^{-NK/12}\,\Theta_\Lambda(2iKt)\prod_{n=1}^\infty P_n\label{ZGaussint}
\end{align}
where, denoting $(K+1)\equiv(1)$, the Gaussian integrals remaining after the commutations of all the oscillators in the products $\mathcal{O}_R^{(k)}\mathcal{O}_L^{(k+1)}$ between ground state operators in (\ref{boseZtrace}) are given by
\begin{align}\label{bosegaussint}
P_n &= \prod_{k=1}^K\int\frac{d^N\mathbf{z}_n^{(k)}d^N\bar{\mathbf{z}}_n^{(k)}d^N\mathbf{w}_n^{(k)}d^N\bar{\mathbf{w}}_n^{(k)}}{\pi^{2N}}\,\,e^{-\mathbf{z}^{(k)}_n\cdot\bar{\mathbf{z}}^{(k)}_n -\mathbf{w}^{(k)}_n\cdot\bar{\mathbf{w}}^{(k)}_n+q^n\sum_l\paren{S_{il}z^{(k)}_{ni}w^{(k)}_{nl}+S_{li}\bar{z}^{(k)}_{nl}\bar{w}^{(k)}_{ni}}}\nonumber\\
&\,\,\,\,\,\,\,\,\,\,\,\,\,\,\,\,\,\,\,\,\times e^{\,q^n\sum_{j\neq i}\sum_l\paren{S_{jl}z^{(k+1)}_{nj}w^{(k)}_{nl} +S_{lj}\bar{z}^{(k+1)}_{nl}\bar{w}^{(k)}_{nj}}}
\end{align}
The lattice theta function and the other factors multiplying the Gaussian integrals in (\ref{ZGaussint}) result from the product of the $K$ operators $G'$ inside the trace in (\ref{boseZtrace}). At this point we could perform the Gaussian integrals in (\ref{bosegaussint}) altogether by way of a determinant, but for the sake of simplifying the calculation we first perform each of the $K$ one-dimensional complex Gaussian integrals in the variables $z_{ni}$, $\bar{z}_{ni}$ and $w_{ni}$, $\bar{w}_{ni}$.
After performing these integrals (see appendix \ref{boseintapp}) we have a reduced expression for the Gaussian integrals
\begin{align}\label{intreduce}
P_n &= D_n^K\prod_{k=1}^K \int\frac{d^{N-1}\mathbf{z}_n^{(k)}d^{N-1}\bar{\mathbf{z}}_n^{(k)}d^{N-1}\mathbf{w}_n^{(k)}d^{N-1}\bar{\mathbf{w}}_n^{(k)}}{\pi^{2N-2}}\,\,e^{-\mathbf{z}^{(k)}_n\cdot\bar{\mathbf{z}}^{(k)}_n -\mathbf{w}^{(k)}_n\cdot\bar{\mathbf{w}}^{(k)}_n+\sum_{j,l\neq i}A^{(k)}_{jl}}
\end{align}
where
\begin{align}\label{lintermreduce}
A^{(k)}_{jl} &= q^n\paren{S_{jl}+q^{2n}D_nS_{ii}S_{ji}S_{il}}\paren{z_{nj}^{(k+1)}w_{nl}^{(k)}+\bar{z}^{(k+1)}_{nj}\bar{w}^{(k)}_{nl}}\nonumber\\
&\,\,\,\,\,\,\,\,\,\,+q^{2n}D_n\paren{S_{ji}S_{li}z^{(k+1)}_{nj}\bar{z}^{(k)}_{nl} +S_{ij}S_{il}w^{(k+1)}_{nj}\bar{w}^{(k)}_{nl}}
\end{align}
and $D_n=\paren{1-q^{2n}S_{ii}^2}^{-1}$. Now we switch to the evaluation of the Gaussian integrals through a determinant, which we do by writing (\ref{intreduce}) as a $4(N-1)K$-dimensional real Gaussian integral
\begin{equation}\label{realboseGint}
P_n=D_n^K\int\frac{d^{4(N-1)K}\mathbf{v}}{\pi^{2(N-1)K}}\,e^{-\mathbf{v}\cdot M_k\mathbf{v}}
\end{equation}
Ordering the real variables according to
$$\mathbf{v}=\paren{\text{Re}\,z_{n1}^{(1)},\,\text{Im}\,z_{n1}^{(1)},\ldots,\,\text{Re}\,z_{nN}^{(1)},\text{Im}\,z_{nN}^{(1)},\,\text{Re}\,w_{n1}^{(1)},\,\text{Im}\,w_{n1}^{(1)},\ldots,\,\text{Re}\,z_{n1}^{(2)},\,\text{Im}\,z_{n1}^{(2)},\,\ldots}$$
we find the matrix exponent has the block circulant form
\begin{equation}\label{boseMK}
M_K=\begin{pmatrix} 1_{4N-4} & C^\text{T} & 0 & \cdots & 0 & C \\ C & 1_{4N-4} & C^\text{T} & \cdots & 0 & 0 \\ 0 & C & 1_{4N-4} & \cdots & 0 & 0 \\ \vdots & \vdots & \vdots & \ddots & \vdots & \vdots \\ 0 & 0 & 0 & \cdots & 1_{4N-4} & C^\text{T} \\ C^\text{T} & 0 & 0 & \cdots & C & 1_{4N-4} \end{pmatrix}
\end{equation}
with off-diagonal blocks themselves in $2\times 2$ block form
\begin{equation}\label{boseC}
C=\frac{1}{2}\begin{pmatrix} X\otimes\paren{1_2+\sigma^2} & 2Y\otimes\sigma^3 \\ 0 & Z\otimes\paren{1_2+\sigma^2} \end{pmatrix}
\end{equation}
and the constituent $(N-1)\times(N-1)$ matrices defined in terms of $q$ and $S$ as
\begin{equation}\label{boseXYZ}
X_{jl}=-q^{2n}D_nS_{ji}S_{li}\,,\,\,\,\,\,Y_{jl}=-q^n\left(S_{jl}+q^{2n}D_nS_{ii}S_{ji}S_{il}\right),\,\,\,\,\,\,Z_{jl}=-q^{2n}D_nS_{ij}S_{il}
\end{equation}
The Gaussian integral (\ref{realboseGint}) is then evaluated to give 
\begin{align}
P_n &= D_n^K\paren{\det\,M_K}^{-1/2}\nonumber\\
&=\prod_{k=1}^K\paren{1-q^{2n}}^{N-2}\left[1-2\paren{S_{ii}^2+\paren{1-S_{ii}^2}\cos(2\pi k/K)}q^{2n}+q^{4n}\right]
\end{align}
where the determinant is calculated in appendix \ref{detcalcs}. Comparing the above to (\ref{thetaforms}) and employing the identity
\begin{equation}
\prod_{k=1}^{K-1}\sin\frac{\pi k}{K}=\frac{K}{2^{K-1}}
\end{equation}
we can immediately write down the $K$-sheeted partition function in terms of modular functions
\begin{equation}\label{NbosZ}
Z(K)=\text{vol}(\Lambda)^KK\,\mathcal{T}_i^{\,(K-1)/2}\,\Theta_\Lambda(2iKt)\left[\eta\paren{2it}\right]^{-K(N-3)-3}\prod_{k=1}^{K-1}\theta^{-1}_1(\nu_k|2it)
\end{equation}
with
\begin{equation}\label{nuk}
\sin\pi\nu_k=\sqrt{\mathcal{T}_i}\,\sin\frac{\pi k}{K}
\end{equation}
This partition function matches the $N=2$ case (\ref{newZK}), and the oscillator part remains the same for all $N$. Performing an $S$-transformation on (\ref{NbosZ}) yields
\begin{equation}
Z(K) = K^{-(N-2)/2}\paren{\mathcal{T}_i\,\text{vol}(\Lambda)^2}^{(K-1)/2}\paren{2t}^{(K-1)(N-2)/2}e^{\pi [K(N-3)+3]/24t}e^{\varphi(K)/t}+\cdots
\end{equation}
where
\begin{equation}
\varphi(K)=\frac{\pi}{2}\sum_{k=1}^{K-1}\paren{\nu_k-\frac{1}{2}}^2
\end{equation}
and the dots indicate terms that go to zero as $t\rightarrow 0$, corresponding to the removal of the cutoffs. Performing the analytic continuation (reviewed in appendix \ref{bernpolys}) and calculating the derivatives in (\ref{SEEfromZK}), the entanglement entropy is
\begin{equation}\label{boseEE}
\mathcal{S}_i=\frac{1}{2}\,\sigma\big(\sqrt{\mathcal{T}_i}\,\big)\log\frac{L}{\epsilon}+\frac{1}{2}\paren{N-2}[1-\log(2t)]-\frac{1}{2}\log\paren{\mathcal{T}_i\,\text{vol}(\Lambda)^2}
\end{equation}
The universal term in the above has the same functional form regardless of the value of $N$, following exactly the behavior described in (\ref{Calacon}). Also independent of $N$, the constant term retains the same dependence on the physical quantities of the junction. The only explicit dependence on the number of theories in the junction comes in the form of a new term that vanishes when $N=2$, which contains a  subleading $\log(\log(L/\epsilon))$ term, the appearance of such a term in related contexts  has been remarked previously in the literature  \cite{Holzhey:1994we,Donnelly:2015hxa,Michel:2016fex}.  Its presence precisely corresponds to the cases where the central charge differs between the inside and outside of the entangling region in the partially folded picture, and thus not covered in the scope of (\ref{basicEE}). However, as this term does not depend on any of the parameters of the junction it will vanish from all differences in entanglement entropy between different junctions, and thus can be considered unphysical.

\subsection{Fermionic NS junction}

If we try to extend to the general $N$-junction the direct methods used to obtain the fermionic interface entanglement entropy outlined in section \ref{sec3}, we'll need to expand the exponential in the boundary state (\ref{NfermBstate}), unfold the $i$-th fermion, and organize the non-vanishing terms into a $4(N-1)\times 4$ matrix representation of $(I_{1\cdots N,i})_n$. If we then consider the reciprocal entanglement entropy for simplicity, we'll need to calculate the $4\times 4$ matrix representation of the partition function block and find its eigenvalues. It is not clear how these matrix computations can be done for arbitrary $N$. Therefore we will employ the fermionic version of the linearization methods utilized in the bosonic calculation.

We begin with the fermionic analog of (\ref{boslinint}), the complex Grassmann Gaussian integral
\begin{equation}\label{fermlinint}
e^{\mathbf{A}\cdot\mathbf{B}}=\int d^N\eta\,d^N\bar{\eta}\,\,e^{\,\boldsymbol\eta\cdot\bar{\boldsymbol\eta}+\mathbf{A}\cdot\boldsymbol\eta+\bar{\boldsymbol\eta}\cdot\mathbf{B}}
\end{equation}
where $\mathbf{A}$ and $\mathbf{B}$ are now $N$-dimensional vectors of anti-commuting operators, which are taken to be Grassmann-valued, and the measure is defined to be
\begin{equation}
d^N\eta\,d^N\bar{\eta}=d\eta_N\cdots d\eta_1\,d\bar{\eta}_N\cdots d\bar{\eta}_1=(-1)^N d\eta_1\,d\bar{\eta}_1\cdots d\eta_N\,d\bar{\eta}_N
\end{equation}
Note that the ordering of the pairs $d\eta_j\,d\bar{\eta}_j$ in the above can be changed without the introduction of additional minus signs. Using (\ref{fermlinint}) we can linearize the Neveu-Schwarz boundary state (\ref{NfermBstate}) and unfold the $i$-th fermion via (\ref{fermunfold}) to obtain explicit interface and anti-interface operators
\begin{align}
I_{1\ldots N,i} &= \prod_{n\in\mathbb{N}-\tfrac{1}{2}}\int d^N\eta_n\,d^N\bar{\eta}_n\,e^{\,\boldsymbol\eta_n\cdot\bar{\boldsymbol\eta}_n+\sum_{j\neq i}\psi^j_{-n}\eta_{nj}+i\sum_{j\neq i}\sum_lS_{lj}\bar{\eta}_{nl}\bar{\psi}^j_{-n}}\Big(\bigotimes_{j\neq i}|0\rangle\Big)\nonumber\\
&\,\,\,\,\,\,\,\,\,\,\,\,\,\,\,\,\,\,\,\,\otimes\langle 0|\prod_{n\in\mathbb{N}-\tfrac{1}{2}}e^{-i\bar{\psi}^i_n\eta_{ni}-\sum_lS_{li}\bar{\eta}_{nl}\psi^i_n}\\
(I_{1\ldots N,i})^\dagger &= \prod_{n\in\mathbb{N}-\tfrac{1}{2}}\int d^N\chi_n\,d^N\bar{\chi}_n\,e^{\,\boldsymbol\chi_n\cdot\bar{\boldsymbol\chi}_n+\sum_lS_{il}\bar{\psi}^i_{-n}\chi_{nl}+i\bar{\chi}_{ni}\psi^i_{-n}}|0\rangle\nonumber\\
&\,\,\,\,\,\,\,\,\,\,\,\,\,\,\,\,\,\,\,\,\otimes\Big(\bigotimes_{j\neq i}\langle 0|\Big)\prod_{n\in\mathbb{N}-\tfrac{1}{2}}e^{\,i\sum_{j\neq i}\sum_lS_{jl}\psi^j_n\chi_{nl}+\sum_{j\neq i}\bar{\chi}_{nj}\bar{\psi}^j_n}
\end{align}
With these expressions we can calculate the commutations between the various products of Grassmann variables and Grassmann-valued operators appearing in (\ref{ZKNdef}) in terms of the operator anti-commutators, e.g. for $\{\alpha,\beta\}=\{\beta,\theta\}=\{\alpha,\phi\}=0$ it follows that
\begin{equation}
[\alpha\theta,\beta\phi]=-\alpha\beta\{\theta,\phi\}
\end{equation}
All that remains in order to calculate the NS partition function block
\begin{equation}
J = q^{-N/24}I_{1\ldots N,i}\,q^{L^i_0+\bar{L}^i_0}\paren{I_{1\ldots N,i}}^\dagger q^{\,\sum_{j\neq i} (L^j_0+\bar{L}^j_0)}
\end{equation}
is for a fermionic version of the identities in (\ref{boseViraident}) to hold. Expanding
\begin{align}
q^{n\psi_{-n}\psi_n} &= \sum_{m=0}^\infty\frac{1}{m!}\paren{n\log q}^m\paren{\psi_{-n}\psi_n}^m\nonumber\\
&= 1+\sum_{m=1}^\infty\frac{1}{m!}\paren{n\log q}^m\psi_{-n}\psi_n=1+\paren{q^n-1}\psi_{-n}\psi_n
\end{align}
we can explicitly expand and recombine the product
\begin{align}
e^{\beta\psi_n}q^{n\psi_{-n}\psi_n} &= \paren{1+\beta\psi_n}\paren{1+\paren{q^n-1}\psi_{-n}\psi_n}\nonumber\\
&= 1+q^n\beta\psi_n+\paren{q^n-1}\psi_{-n}\psi_n=q^{n\psi_{-n}\psi_n}e^{q^n\beta\psi_n}
\end{align}
which shows that indeed
\begin{equation}\label{fermViraident}
e^{\beta \psi_n}q^{L_0} = q^{L_0}e^{q^n\beta \psi_n}\,\,\,\,\,\,\text{and}\,\,\,\,\,\,e^{\beta\bar{\psi}_n}q^{\bar{L}_0} = q^{\bar{L}_0}e^{q^n\beta\bar{\psi}_n}
\end{equation}
exactly as in the bosonic case. Performing the commutator calculations between the exponentials of the oscillators of the $i$-th fermion, in a similar manner to those behind (\ref{boseJexp}), we obtain
\begin{equation}\label{fermJ}
J = \prod_{n\in\mathbb{N}-\tfrac{1}{2}}\int d^N\eta_n\,d^N\bar{\eta}_n\,d^N\chi_n\,d^N\bar{\chi}_n\,\,e^{\,\boldsymbol\eta_n\cdot\bar{\boldsymbol\eta}_n +\boldsymbol\chi_n\cdot\bar{\boldsymbol\chi}_n}\,e^{\,iq^n\sum_j\paren{S_{ij}\eta_{ni}\chi_{nj} +S_{ji}\bar{\eta}_{nj}\bar{\chi}_{ni}}}\,\mathcal{O}_LG'\mathcal{O}_R
\end{equation}
where the remaining oscillators are contained in
\begin{align}
\mathcal{O}_L &= \prod_{n\in\mathbb{N}-\tfrac{1}{2}}\exp\Big[\sum_{j\neq i}\psi^j_{-n}\eta_{nj}+i\sum_l\sum_{j\neq i}S_{lj}\bar{\eta}_{nl}\bar{\psi}^j_{-n}\Big]\\
\mathcal{O}_R &= \prod_{n\in\mathbb{N}-\tfrac{1}{2}}\exp\Big[q^n\Big(\sum_{j\neq i}\bar{\chi}_{nj}\bar{\psi}^j_{n}+i\sum_l\sum_{j\neq i}S_{jl}\psi^j_{n}\chi_{nl}\Big)\Big]
\end{align}
with ground state operator
\begin{equation}
G' = q^{-N/24}\Big(\bigotimes_{j\neq i}|0\rangle\Big)\otimes\Big(\bigotimes_{j\neq i}\langle 0|\Big)
\end{equation}
We can now write the $K$-sheeted partition function (\ref{ZKNdef}) in terms of the block (\ref{fermJ}) as
\begin{align}
Z(K) &= \text{Tr}_{1\ldots N}\paren{J^K}\\
&= \prod_{n\in\mathbb{N}-\tfrac{1}{2}}\int\prod_{k=1}^K d^N\eta_n^{(k)}d^N\bar{\eta}_n^{(k)}d^N\chi_n^{(k)}d^N\bar{\chi}_n^{(k)}\,e^{\,\boldsymbol\eta^{(k)}_n\cdot\bar{\boldsymbol\eta}^{(k)}_n +\boldsymbol\chi^{(k)}_n\cdot\bar{\boldsymbol\chi}^{(k)}_n+iq^n\sum_j\paren{S_{ij}\eta^{(k)}_{ni}\chi^{(k)}_{nj} +S_{ji}\bar{\eta}^{(k)}_{nj}\bar{\chi}^{(k)}_{ni}}}\nonumber\\
&\,\,\,\,\,\,\,\,\,\,\,\,\,\,\,\,\,\,\,\,\times\text{Tr}_{1\ldots N}\paren{G'\mathcal{O}^{(1)}_R\mathcal{O}^{(2)}_LG'\mathcal{O}^{(2)}_R\cdots \mathcal{O}^{(K)}_LG'\mathcal{O}^{(K)}_R\mathcal{O}^{(1)}_L}\label{fermZKtrace}\\
&=q^{-NK/24}\prod_{n\in\mathbb{N}-\tfrac{1}{2}} P_n\label{ZKfermint}
\end{align}
where, denoting $(K+1)\equiv(1)$, the Gaussian integrals remaining after all the commutations of all the oscillators in the products $\mathcal{O}_R^{(k)}\mathcal{O}_L^{(k+1)}$ between vacuum states in (\ref{fermZKtrace}) are given by
\begin{align}\label{fermgaussint}
P_n &= \prod_{k=1}^K\int d^N\eta_n^{(k)}d^N\bar{\eta}_n^{(k)}d^N\chi_n^{(k)}d^N\bar{\chi}_n^{(k)}\,e^{\,\boldsymbol\eta^{(k)}_n\cdot\bar{\boldsymbol\eta}^{(k)}_n +\boldsymbol\chi^{(k)}_n\cdot\bar{\boldsymbol\chi}^{(k)}_n+iq^n\sum_j\paren{S_{ij}\eta^{(k)}_{ni}\chi^{(k)}_{nj} +S_{ji}\bar{\eta}^{(k)}_{nj}\bar{\chi}^{(k)}_{ni}}}\nonumber\\
&\,\,\,\,\,\,\,\,\,\,\,\,\,\,\,\,\,\,\,\,\times e^{\,iq^n\sum_l\sum_{j\neq i}\paren{S_{jl}\eta^{(k+1)}_{nj}\chi^{(k)}_{nl} +S_{lj}\bar{\eta}^{(k+1)}_{nl}\bar{\chi}^{(k)}_{nj}}}
\end{align}
At this point we could perform the integrals in (\ref{fermgaussint}) altogether by way of a determinant, but for the sake of simplifying the calculation we first perform each of the $K$ one-dimensional complex Grassmann Gaussian integrals in the variables $\eta_{ni}$, $\bar{\eta}_{ni}$ and $\chi_{ni}$, $\bar{\chi}_{ni}$. After performing these integrals (see appendix \ref{fermintapp}) we have a reduced expression for the Gaussian integrals
\begin{align}\label{fermgaussintred}
P_n &= D_n^{-K}\prod_{k=1}^K \int d^{N-1}\eta_n^{(k)}d^{N-1}\bar{\eta}_n^{(k)}d^{N-1}\chi_n^{(k)}d^{N-1}\bar{\chi}_n^{(k)}\,e^{\,\boldsymbol\eta^{(k)}_n\cdot\bar{\boldsymbol\eta}^{(k)}_n +\boldsymbol\chi^{(k)}_n\cdot\bar{\boldsymbol\chi}^{(k)}_n+\sum_{j,l\neq i}A^{(k)}_{jl}}
\end{align}
where
\begin{align}\label{fermgaussintexp}
A^{(k)}_{jl} &= iq^n\paren{S_{jl}-q^{2n}D_nS_{ii}S_{ji}S_{il}}\paren{\eta_{nj}^{(k+1)}\chi_{nl}^{(k)}+\bar{\eta}^{(k+1)}_{nj}\bar{\chi}^{(k)}_{nl}}\nonumber\\
&\,\,\,\,\,\,\,\,\,\,+q^{2n}D_n\paren{S_{ji}S_{li}\eta^{(k+1)}_{nj}\bar{\eta}^{(k)}_{nl} +S_{ij}S_{il}\chi^{(k+1)}_{nj}\bar{\chi}^{(k)}_{nl}}
\end{align}
and $D_n=(1+q^{2n}S^2_{ii})^{-1}$. Now we switch to the evaluation of the Gaussian integrals through a determinant, which we do by writing (\ref{fermgaussintred}) as a $4(N-1)K$-dimensional real Grassmann Gaussian integral
\begin{equation}\label{fermbigint}
P_n=D_n^{-K}\paren{-1}^{(N-1)K}\int d^{\,4(N-1)K}\theta\,e^{\,\tfrac{1}{2}\boldsymbol\theta\cdot M_k\boldsymbol\theta}
\end{equation}
Ordering the real Grassmann variables according to
$$\boldsymbol\theta=\paren{\text{Re}\,\eta_{n1}^{(1)},\,\text{Im}\,\eta_{n1}^{(1)},\ldots,\,\text{Re}\,\eta_{nN}^{(1)},\text{Im}\,\eta_{nN}^{(1)},\,\text{Re}\,\chi_{n1}^{(1)},\,\text{Im}\,\chi_{n1}^{(1)},\ldots,\,\text{Re}\,\eta_{n1}^{(2)},\,\text{Im}\,\eta_{n1}^{(2)},\,\ldots}$$
we find the matrix exponent has the block circulant form
\begin{equation}
M_K=\begin{pmatrix} 1_{2N-2}\otimes\sigma^2 & -C^\text{T} & 0 & \cdots & 0 & C \\ C & 1_{2N-2}\otimes\sigma^2 & -C^\text{T} & \cdots & 0 & 0 \\ 0 & C & 1_{2N-2}\otimes\sigma^2 & \cdots & 0 & 0 \\ \vdots & \vdots & \vdots & \ddots & \vdots & \vdots \\ 0 & 0 & 0 & \cdots & 1_{2N-2}\otimes\sigma^2 & -C^\text{T} \\ -C^\text{T} & 0 & 0 & \cdots & C & 1_{2N-2}\otimes\sigma^2 \end{pmatrix}
\end{equation}
with off-diagonal blocks themselves in $2\times 2$ block form
\begin{equation}
C=\frac{1}{2}\begin{pmatrix} X\otimes\paren{1_2+\sigma^2} & 2Y\otimes\sigma^3 \\ 0 & Z\otimes\paren{1_2+\sigma^2} \end{pmatrix}
\end{equation}
where the matrices $X$, $Y$, and $Z$ are the same as the bosonic case (\ref{boseXYZ}) only with the replacement $q^n\rightarrow -iq^n$. The Gaussian integral (\ref{fermbigint}) is then evaluated to give 
\begin{align}
P_n&=D_n^{-K}\paren{-1}^{(N-1)K}\paren{\det\,M_K}^{1/2}\nonumber\\
&= \prod_{k=1}^K\paren{1+q^{2n}}^{N-2}\left[1+2\paren{S_{ii}^2+\paren{1-S_{ii}^2}\cos(2\pi k/K)}q^{2n}+q^{4n}\right]
\end{align}
where the determinant is calculated in appendix \ref{detcalcs}. With this final expression for the integrals, we are able to write the replicated NS partition function in terms of modular functions and make an $S$-transformation
\begin{align}
Z(K) &= \left[\eta\paren{2it}\right]^{-NK/2}\left[\theta_3(2it)\right]^{K(N-2)/2+1}\prod_{k=1}^{K-1}\theta_3(\nu_k|2it)\\
&= e^{\pi NK/48t}e^{-\vartheta(K)/t}+\cdots
\end{align}
where $\nu_k$ is given by (\ref{nuk}), the exponent $\vartheta(K)$ is
\begin{equation}\label{fermKexp}
\vartheta(K)=\frac{\pi}{2}\sum_{k=1}^{K-1}\nu_k^2
\end{equation}
and the dots indicate terms which vanish as $t\rightarrow 0$. The entanglement entropy is then
\begin{equation}\label{fermEE}
\mathcal{S}_i=\frac{1}{2}\,\left[\frac{1}{2}\sqrt{\mathcal{T}_i}-\sigma\big(\sqrt{\mathcal{T}_i}\,\big)\right]\log\frac{L}{\epsilon}
\end{equation}
after analytically continuing (\ref{fermKexp}), see the review in appendix \ref{bernpolys} for details, and taking the derivatives in (\ref{SEEfromZK}). As in the bosonic case, the entanglement entropy (\ref{fermEE}) shows the same $N$-independent behavior described in (\ref{Calacon}).

\subsection{BPS junction}

Until this point we have been considering interfaces and junctions that preserve conformal symmetry, i.e. satisfy (\ref{intcond}) in the unfolded or partially folded picture. Since we have been working with free conformal bosons and fermions we could further consider interfaces and junctions that also preserve supersymmetry.

Whereas the conformal condition (\ref{intcond}) enforces continuity of the stress tensor across the interface, if we further require continuity of the supercurrent the interface operator must satisfy
\begin{equation}
\paren{G^1_n-i\eta^1\bar{G}^1_{-n}}I_{1,2}=I_{1,2}\paren{G^2_n-i\eta^2\bar{G}^2_{-n}}
\end{equation}
with supercurrent modes
\begin{equation}
G^i_n=\sum_{m=-\infty}^\infty a^i_{-m}\psi^i_{n+m}\,,\,\,\,\,\,\,\bar{G}^i_n=\sum_{m=-\infty}^\infty \bar{a}^i_{-m}\bar{\psi}^i_{n+m}
\end{equation}
The constants $\eta^1=\pm 1$ and $\eta^2=\pm 1$ determine the type of supersymmetry in CFT$_1$ and CFT$_2$, respectively, and do not need to be equal. The generalization to a partially folded $N$-junction is
\begin{equation}\label{superjuncon}
\sum_{j\neq i}\paren{G^j_n-i\eta^j\bar{G}^j_{-n}}I_{1\cdots N,i}=I_{1\cdots N,i}\paren{G^i_n-i\eta^i\bar{G}^i_{-n}}
\end{equation}
If $\eta^j=1$ for all $j=1,\ldots,N$ then the operator produced by unfolding the supersymmetric boundary state
\begin{equation}\label{superbndstate}
|S\rangle\rangle_\text{super}=|S\rangle\rangle_\text{bos}\otimes|S\rangle\rangle_\text{NS}
\end{equation}
will satisfy (\ref{superjuncon}). Furthermore, if we redefine $\bar{\psi}^j\rightarrow\eta^j\bar{\psi}^j$ then the $\eta^j$ are absorbed into the interface operator through $S_{ij}\rightarrow S'_{ij}=\eta^jS_{ij}$. Introducing these factors does not change the entropy calculations, as $S'$ is still an element of $O(N)$ and ${S'_{ii}}^2=S_{ii}^2$ regardless of the values of the $\eta^j$. Thus for the purposes of calculating the entanglement entropy we proceed as though the supersymmetric boundary state (\ref{superbndstate}) unfolds simply into a supersymmetry-preserving interface operator no matter the types of supersymmetry present in the individual CFTs. The replicated partition function is then the product
\begin{equation}
Z_\text{super}(K)=Z_\text{bos}(K)Z_\text{NS}(K)
\end{equation}
and through the logarithm the entanglement entropy is the sum
\begin{equation}\label{superEE}
\mathcal{S}_\text{super}=\mathcal{S}_\text{bos}+\mathcal{S}_\text{NS}=\frac{1}{4}\sqrt{\mathcal{T}_i}\,\log\frac{L}{\epsilon}+\frac{1}{2}\paren{N-2}[1-\log(2t)]-\frac{1}{2}\log\paren{\mathcal{T}_i\,\text{vol}(\Lambda)^2}
\end{equation}
This simplification of the oscillator contribution to the universal term of the entanglement entropy is precisely the same as in \cite{Brehm:2015lja} for $N=2$.

\section{Specific 3-junction geometries}
\label{sec5}

We now focus on constructing the explicit boundary states describing bosonic $3$-junctions using similar methods to those used to construct (\ref{D1theta}) and (\ref{2D2D0}). We will also relate the quantities relevant to the entanglement entropy, the total transmission $\mathcal{T}_i$ and unit cell volume $\text{vol}(\Lambda)$, to the geometry of the corresponding D-branes describing the junctions in the folded picture.

\subsection{Boundary state construction}
Following the procedure outlined in \cite{Bachas:2007td}, we begin with the boundary state 
\begin{equation}
|k_2\text{D}2/k_1\text{D}0,0,0\rangle\rangle=|k_2\text{D}2/k_1\text{D}0\rangle\rangle\otimes |\text{D}0\rangle\rangle
\end{equation}
corresponding to $k_2$ D2-branes in the $\varphi^1\varphi^2$-plane bound to $k_1$ D0-branes, which we rotate to an arbitrary orientation in the compactification lattice. Through translation we can specify an arbitrary orientation by the axis intercepts $q_1R_1\,\hat{\varphi}_1$, $q_2R_2\,\hat{\varphi}_2$, and $q_3R_3\,\hat{\varphi}_3$. Such a plane will have an area vector equal to
\begin{equation}\label{normvec}
\mathbf{A}=q_2q_3R_2R_3\,\hat{\varphi}_1+q_1q_3R_1R_3\,\hat{\varphi}_2+q_1q_2R_1R_2\,\hat{\varphi}_3
\end{equation}
and thus the rotation transformation needed will be $\mathcal{R}(\theta,\phi)=\mathcal{R}_3(\phi)\mathcal{R}_2(\theta)$ where 
\begin{equation}
\tan\theta=\frac{q_1q_2R_1R_2}{\sqrt{\paren{q_2q_3R_2R_3}^{2}+\paren{q_1q_3R_1R_3}^{2}}}\,,\,\,\,\,\,\,\tan\phi=\frac{q_1R_1}{q_2R_2}
\end{equation}
in order to obtain the rotated D-brane state $|k_2\text{D}2/k_1\text{D}0,\theta(q_1,q_2,q_3),\phi(q_1,q_2)\rangle\rangle$.
\begin{figure}[t]
\centering
\includegraphics[scale=1.5]{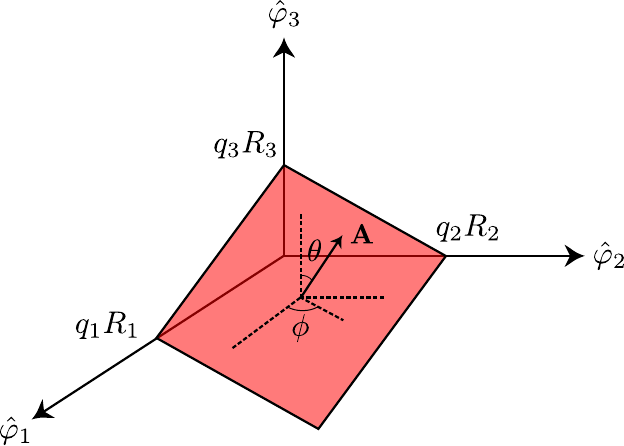}
\caption{A D2-brane wrapping the bosonic 3-torus continued into the compactification lattice so as to show the axis intercepts $q_iR_i\,\hat{\varphi}_i$; see figure \ref{branecell} for the unit cell wrapping for a specific case. The polar and azimuthal angles that specify the rotation that takes the D2-brane in the $\varphi^1\varphi^2$-plane into this pictured D2-brane are also shown.}
\end{figure}
\begin{figure}
\centering
\includegraphics[scale=1.0]{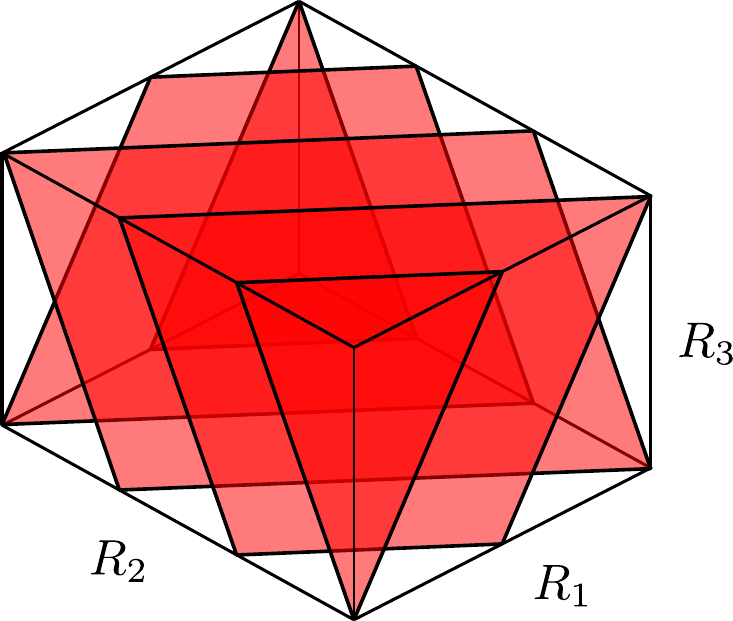}
\caption{A D2-brane wrapping the bosonic 3-torus shown in the unit cell of the compactification lattice. The above corresponds to the parameters $q_1=3$, $q_2=2$, and $q_3=6$.\label{branecell}}
\end{figure}
To do this we will transform the boundary conditions
\begin{equation}\label{unrotbcs}
\left[g_{ij}\paren{a_n^j+\bar{a}_{-n}^j}+b_{ij}\paren{a_n^j-\bar{a}_{-n}^j}+\delta_{i3}\delta_{3j}\paren{a_n^j-\bar{a}_{-n}^j}\right]|k_2\text{D}2/k_1\text{D}0,0,0\rangle\rangle=0
\end{equation}
where $n\geq 0$ and 
\begin{equation}
g=\begin{pmatrix} 1 & 0 & 0 \\ 0 & 1 & 0 \\ 0 & 0 & 0 \end{pmatrix}\,,\,\,\,\,\,b=\frac{k_1\alpha'}{k_2R_1R_2}\begin{pmatrix} 0 & -1 & 0 \\ 1 & 0 & 0 \\ 0 & 0 & 0 \end{pmatrix}
\end{equation}
The metric $g$ and $(E_{33})_{ij}\equiv\delta_{i3}\delta_{3j}$ will simply transform by similarity; however, the magnetic field will undergo an angle-dependent scaling in addition to the rotation in order for the boundary state to correspond to a bound state between $k_2$ D2-branes and $k_1$ D0-branes at all angles. Explicitly, the transformation of the magnetic field is determined through two conditions: (1) the magnetic field is oriented along the $-\hat{\mathbf{A}}$ direction; that is, perpendicular to the D2-branes
\begin{equation}
b_{ij}(\theta,\phi)=\beta(\theta,\phi)\,\varepsilon_{ijk}\mathcal{R}^{k3}(\theta,\phi)
\end{equation}
and (2) the Dirac quantization condition is met at all angles
\begin{equation}
k_2\int_{\text{D2}} F=-k_1\alpha'\,\,\,\,\,\text{with}\,\,\,\,\,F=\frac{1}{2}\,b_{ij}\,d\varphi^i\wedge d\varphi^j
\end{equation}
Enforcing these conditions gives
\begin{equation}\label{corrb}
b_{ij}(\theta,\phi)=\frac{-k_1\alpha'\varepsilon_{ijk}\mathcal{R}^{k3}(\theta,\phi)}{k_2(q_1q_2R_1R_2\cos\theta+q_3R_3\sin\theta\paren{q_1R_1\sin\phi+q_2R_2\cos\phi})}
\end{equation}
The exponent of the rotated state is then found from the boundary conditions
\begin{equation}\label{sfrombcs}
\paren{M_{ij}a^j_n+\bar{M}_{ij}\bar{a}^j_{-n}}|S\rangle\rangle=0\,\,\Longrightarrow\,\,S=M^{-1}\bar{M}
\end{equation}
so that after transforming (\ref{unrotbcs}) we have from (\ref{sfrombcs}) that
\begin{equation}\label{bigsdef}
S(\theta,\phi) = \paren{1_3+b(\theta,\phi)}^{-1}\left[b(\theta,\phi) +\mathcal{R}(\theta,\phi)\paren{E_{33}-g}\mathcal{R}^\text{T}(\theta,\phi)\right]
\end{equation}
where $b$ is given by (\ref{corrb}). It is important to note that $S$ in (\ref{bigsdef}) is a (special) orthogonal matrix.

The next step in our construction will be to find all zero modes that are consistent with (\ref{bigsdef}). These admissible zero modes
\begin{equation}\label{zeromode}
\bigotimes_{i=1}^3|n_i,w_i\rangle
\end{equation}
are determined by the $n=0$ rotated version of (\ref{unrotbcs}), which upon acting on (\ref{zeromode}) reduce to
\begin{align}\label{boundcondrot}
\frac{q_1R_1}{k_2A^2}&\left[q_3R_3^2\paren{k_1w_3+k_2q_3\paren{q_1n_1-q_2n_2}}-q_2R_2^2\paren{k_1w_2+k_2q_2\paren{q_3n_3-q_1n_1}}\right]\nonumber\\
&\,\,\,\,\,\,\,\,\,\,\,\,\,\,\,\,\,\,\,\,+\frac{q_2q_3V^2}{R_1A^2\alpha'}\paren{q_2q_3w_1+q_1q_3w_2+q_1q_2w_3}=0
\end{align}
and the other two cyclic permutations of the indices, where $V$ is the volume of the 3-torus. The first line of (\ref{boundcondrot}) is the contribution to the boundary conditions of the D2-branes with magnetic flux, and the second line is the contribution due to zero winding in the direction perpendicular to the D2-branes. Isolating the dependence on the radii we arrive at the winding constraint
\begin{equation}\label{D0ground}
q_2q_3w_1+q_1q_3w_2+q_1q_2w_3 = 0
\end{equation}
and the three constraint equations given by
\begin{equation}\label{D2ground}
k_1w_1+k_2q_1\left[q_2n_2-q_3n_3\right]=0
\end{equation}
and the other two cyclic permutations of the indices. As long as $k_1\neq 0$ and $k_2\neq 0$ then (\ref{D0ground}) is satisfied by any set of winding numbers that satisfy (\ref{D2ground}). The most general solution to (\ref{D2ground}) is given by
\begin{equation}\label{zmgensol}
n_1(\mathbf{m},\gamma) = k_1m_1+q_2q_3\gamma\,,\,\,\,\,\,\,w_1(\mathbf{m})=k_2q_1\paren{q_3m_3-q_2m_2}
\end{equation}
and the other two cyclic permutations of the indices. Since there are four undetermined integers ($m_1$, $m_2$, $m_3$, and $\gamma$) appearing in (\ref{zmgensol}), this general solution does not specify a basis for $\Lambda$ but rather a generating set. Noticing that
\begin{align}
w_i(m_1,m_2,m_3) &= w_i(m_1+q_2q_3\delta,m_2+q_1q_3\delta,m_3+q_1q_2\delta)\\
n_i(m_1,m_2,m_3,\gamma) &= n_i(m_1+q_2q_3\delta,m_2+q_1q_3\delta,m_3+q_1q_2\delta,\gamma-k_1\delta)
\end{align}
for some integer $\delta$, we see that choices of $\gamma$ modulo $k_1$ correspond to distinct translations of the sublattice generated by summation over $\mathbf{m}\in\mathbb{Z}_3$. Thus, the lattice-sum zero mode in (\ref{NbosBstate}) is parametrized as
\begin{equation}
\sum_{\gamma=0}^{k_1-1}\sum_{\mathbf{m}\in\mathbb{Z}_3}e^{i\delta_{\mathbf{m},\gamma}}\bigotimes_{i=1}^3|n_i(\mathbf{m},\gamma),w_i(\mathbf{m})\rangle
\end{equation}
Applying the result (\ref{volfromgen}), we find
\begin{equation}\label{D2detQ}
\text{vol}(\Lambda) = \frac{k_2^2A^2+k_1^2{\alpha'}^2}{{\alpha'}^2V\sqrt{(2/\alpha')^3}}
\end{equation}
It is known \cite{Harvey:1999gq} that the boundary entropy $g=\langle 0|S\rangle\rangle$ for a pure D$p$-brane in the bosonic $N$-torus is of the form
\begin{equation}
g^2_{\text{D}p}=\frac{V_p^2}{{\alpha'}^{\,p}V_{T^N}\sqrt{(2/\alpha')^N}}
\end{equation}
which gives the suggestive form
\begin{equation}
\text{vol}(\Lambda)=k_2^2g_{\text{D}2}^2+k_1^2g_{\text{D}0}^2
\end{equation}

If any of $q_1$, $q_2$, $q_3$, $k_1$, or $k_2$ are zero then the constraints of (\ref{D2ground}) are relaxed and (\ref{D0ground}) needs to be considered as well, so that (\ref{zmgensol}) no longer represents all admissible zero modes. However, $\text{vol}(\Lambda)$ remains of the same form as (\ref{D2detQ}) in each case. For example, if $q_1=0$ ($q_2=q_3=1$) then
\begin{equation}\label{Omegaq10}
\bigotimes_{i=1}^3|n_i,w_i\rangle = |m_1,0\rangle\otimes|k_1m_2,-k_2m_3\rangle\otimes|k_1m_3,k_2m_2\rangle
\end{equation}
which corresponds precisely to the factorizable state $|\text{D}0\rangle\rangle\otimes|k_2\text{D}2/k_1\text{D}0\rangle\rangle$ describing $k_2$ D2-branes bound to $k_1$ D0-branes in the $\varphi^2\varphi^3$-plane. The special case $k_1=0$ and $k_2=1$ corresponds to a rotated pure D2-brane, with the associated boundary conditions solved by
\begin{equation}\label{pureD2zm}
\bigotimes_{i=1}^3|n_i,w_i\rangle = |q_2q_3m_1,-q_1m_2\rangle\otimes|q_1q_3m_1,-q_2m_3\rangle\otimes|q_1q_2m_1,q_3(m_2+m_3)\rangle
\end{equation}
Lastly, the case $k_2=0$ and $k_1=1$ corresponds to a pure D0-brane where the boundary state is $|\text{D}0\rangle\rangle\otimes|\text{D}0\rangle\rangle\otimes|\text{D}0\rangle\rangle$.

The other class of boundary states, the D1/D3 system, are T-dual to those of the D2/D0 system. Performing a T-duality transformation on all of the three bosons maps the boundary state of $k_2$ D2-branes with area vector $\mathbf{A}$ given in (\ref{normvec}) bound to $k_1$ D0-branes onto the boundary state of $k_2$ D1-branes with length vector
\begin{equation}
\boldsymbol\ell=q_2q_3R_1\,\hat{\varphi}_1+q_1q_3R_2\,\hat{\varphi}_2+q_1q_2R_3\,\hat{\varphi}_3
\end{equation}
bound to $k_1$ D3-branes. Applying the T-duality transformation rules (\ref{Tdualrules}), the matrix exponent of this second class of boundary states is found from (\ref{bigsdef}) to be
\begin{equation}\label{dualsdef}
S'(\theta',\phi') = -\paren{1_3+b'(\theta',\phi')}^{-1}\left[b(\theta',\phi') +\mathcal{R}(\theta',\phi')\paren{E_{33}-g}\mathcal{R}^\text{T}(\theta',\phi')\right]
\end{equation}
for magnetic field
\begin{equation}
b'_{ij}(\theta',\phi')=\frac{-k_1V\varepsilon_{ijk}\mathcal{R}^{k3}(\theta',\phi')}{k_2\alpha'(q_1q_2R_3\cos\theta'+q_3\sin\theta'\paren{q_1R_2\sin\phi'+q_2R_1\cos\phi'})}
\end{equation}
and angles
\begin{equation}
\tan\theta'=\frac{q_1q_2R_3}{\sqrt{\paren{q_2q_3R_1}^2+\paren{q_1q_3R_2}^2}}\,,\,\,\,\,\,\,\tan\phi'=\frac{q_1R_2}{q_2R_1}
\end{equation}
The admissible zero modes for all cases considered before for the D2/D0 system are given by (\ref{zmgensol}), (\ref{Omegaq10}), and (\ref{pureD2zm}) with momenta and windings exchanged for each of the bosons. Taking $R_i\rightarrow\alpha'/R_i$ for all $i=1,2,3$ in (\ref{D2detQ}), the volume of the unit cell of $\Lambda'$ is
\begin{equation}\label{dualnorm}
\text{vol}(\Lambda') = \frac{k_1^2V^2+k_2^2\ell^2{\alpha'}^2}{{\alpha'}^3V\sqrt{(2/\alpha')^3}} = k_1^2g^2_{\text{D}3}+k_2^2g^2_{\text{D}1}
\end{equation}

Lastly, there are some boundary states of the D2/D0 system that are not covered by the construction above; namely those where the D2-branes coincide with exactly one of the $\varphi^i$-axes. For these we rotate the boundary state corresponding to $k_2$ D2-branes in the $\varphi^1\varphi^2$-plane bound to $k_1$ D0-branes about the $\varphi^1$-axis, and all other D2/D0 bound states can be found by suitable permutations of the boson indices. For a rotation angle
\begin{equation}
\tan\xi=\frac{p_3R_3}{p_2R_2}
\end{equation}
the D2-branes will have a corresponding area vector
\begin{equation}\label{specialavec}
\mathbf{A} = -p_3R_1R_3\,\hat{\varphi}_2+p_2R_1R_2\,\hat{\varphi}_3
\end{equation}
with a matrix exponent
\begin{equation}\label{specialSdef}
S(\xi) = \paren{1_3+b(\xi)}^{-1}\left[b(\xi) +\mathcal{R}_1(\xi)\paren{E_{33}-g}\mathcal{R}_1^\text{T}(\xi)\right]
\end{equation}
where the magnetic field is given by
\begin{equation}
b_{ij}(\xi)=\frac{-k_1\alpha'\varepsilon_{ijk}\mathcal{R}^{k3}_1(\xi)}{k_2R_1(p_2R_2\cos\xi+p_3R_3\sin\xi)}
\end{equation}
The admissible zero modes for this boundary state are
\begin{equation}\label{specialzm}
\bigotimes_{i=1}^3|n_i,w_i\rangle = |k_1m_1,k_2(p_2m_2+p_3m_3)\rangle\otimes| -k_1m_2,k_2p_2m_1\rangle\otimes| -k_1m_3,k_2p_3m_1\rangle
\end{equation}
producing a normalization factor of the same form as (\ref{D2detQ}) for the area vector (\ref{specialavec}). Following again the transformation rules in (\ref{Tdualrules}), the dual D1/D3 bound state has a length vector
\begin{equation}\label{specialD1D3}
\boldsymbol\ell = -p_3R_2\,\hat{\varphi}_2+p_2R_3\,\hat{\varphi}_3
\end{equation}
for the D1-branes, which is a rotation about the $\varphi^1$-axis of the bound state with D1-branes along the $\varphi^3$-axis by an angle
\begin{equation}
\tan\xi'=\frac{p_3R_2}{p_2R_3}
\end{equation}
The matrix exponent is then determined from (\ref{specialSdef}) to be
\begin{equation}\label{sdsdef}
S'(\xi') = -\paren{1_3+b'(\xi')}^{-1}\left[b'(\xi') +\mathcal{R}_1(\xi')\paren{E_{33}-g}\mathcal{R}_1^\text{T}(\xi')\right]
\end{equation}
where the magnetic field is given by
\begin{equation}
b'_{ij}(\xi')=\frac{-k_1V\varepsilon_{ijk}\mathcal{R}^{k3}_1(\xi')}{k_2\alpha'(p_2R_3\cos\xi'+p_3R_2\sin\xi')}
\end{equation}
The admissible zero modes are (\ref{specialzm}) with the momenta and windings exchanged for each of the bosons, producing a normalization factor of the same form as (\ref{dualnorm}) for the length vector (\ref{specialD1D3}).

\subsection{Transmission and entanglement entropy}

With the normalization factors (\ref{D2detQ}) and (\ref{dualnorm}) the only other physical quantity remaining in the entanglement entropy (\ref{boseEE}) is the total transmission $\mathcal{T}_i$ of the $i$-th boson. From the matrix exponents (\ref{bigsdef}) and (\ref{sdsdef}), the transmission coefficients of the D2/D0 system are expressed in terms of the area vector of the D2-branes as
\begin{equation}\label{3bostrans}
\mathcal{T}_i = \frac{4k_2^2(A^2-A_{i}^2)(k_2^2A_{i}^2+k_1^2{\alpha'}^2)}{(k_2^2A^2+k_1^2{\alpha'}^2)^2}
\end{equation}
where $A_{i}=\mathbf{A}\cdot\hat{\varphi}_i$ is the area of each of the D2-branes projected onto the plane with normal $\hat{\varphi}_i$. For the D1/D3 system the transmission coefficients obtained from (\ref{3bostrans}) by T-duality are expressed in terms of the length vector of the D1-branes as
\begin{equation}\label{3bostransT}
\mathcal{T}_i=\frac{4k_2^2{\alpha'}^2\paren{\ell^2-\ell^2_{i}}\paren{k_1^2V^2+k_2^2\ell^2_{i}{\alpha'}^2}}{\paren{k_1^2V^2+k_2^2\ell^2{\alpha'}^2}^2}
\end{equation}
where $\ell_i=\boldsymbol\ell\cdot\hat{\varphi}_i$ is the projected length of each of the D1-branes along $\hat{\varphi}_i$. At this point we have found all the boundary states describing 3-junctions and their physical quantities relevant to the entanglement entropy.

From the form of (\ref{3bostrans}) and (\ref{3bostransT}) the $i$-th boson is seen to decouple either in the case of a pure D0 or D3-brane, or when the area or length vector aligns with the $\varphi^i$-axis. Furthermore, we see that perfectly transmissive junctions (with respect to CFT$_i$) are those where 
\begin{equation}\label{ctranscon}
\frac{A_i^2}{A^2}=\frac{1}{2}-\frac{1}{2}\paren{\frac{k_1\alpha'}{k_2A}}^2\,\,\,\,\,\,\,\,\,\,\text{or}\,\,\,\,\,\,\,\,\,\,\frac{\ell_i^2}{\ell^2}=\frac{1}{2}-\frac{1}{2}\paren{\frac{k_1V}{k_2\ell\alpha'}}^2
\end{equation}
These conditions cannot necessarily be met for general real radii $R_i$ and coupling $\alpha'$, solutions are only possible when ratios of these real numbers are rational. The conditions simplify in the purely geometric cases ($k_1=0$), which are met by D1-branes and D2-branes whose length and area vectors lie on any of the right angle cones about each of the $\varphi^i$-axes. From the form of (\ref{ctranscon}) we see that a completely transmissive junction, $\mathcal{T}_i=1$ for $i=1,2,3$, can only occur when $k_1\neq 0$, $k_2\neq 0$, and the quantities
\begin{align}
\frac{k_1\alpha'}{k_2R_iR_j}\,\,\,\,\,\,\,\,\,\,\text{or}\,\,\,\,\,\,\,\,\,\,\frac{k_1R_iR_j}{k_2\alpha'}
\end{align}
are all integers. The volume of the unit cell reduces to
\begin{equation}
\text{vol}(\Lambda)=\frac{k_1^2}{V}\,\sqrt{2{\alpha'}^3}\,\,\,\,\,\,\,\,\,\,\text{or}\,\,\,\,\,\,\,\,\,\,\text{vol}(\Lambda')=k_1^2V\sqrt{\frac{2}{{\alpha'}^3}}
\end{equation}
in these cases. This result is interesting, as the only the number of D-branes present in the bound state enter into the entanglement entropy of the completely transmissive junctions.

Finally when any of the boundary states align entirely with a single plane, the entanglement entropy reduces to the $N=2$ results with an additional constant term corresponding to the perpendicular factor of the decoupled boson. For example, for (\ref{specialD1D3}) with $k_1=0$ and $k_2=1$ we have
\begin{equation}
\mathcal{T}_3=\sin^22\xi'\,\,\,\,\,\,\,\,\,\,\text{and}\,\,\,\,\,\,\,\,\,\,\mathcal{T}_3\,\text{vol}(\Lambda')^2=p_2^2p_3^2\,\frac{\alpha'}{2R_1^2}
\end{equation}
which differs from (\ref{bose2EE}) only in the additional constant boundary entropy of the Dirichlet boundary condition along the $\hat{\varphi}_1$ direction.

\section{Discussion}
\label{sec6}

The main new results are  the generalization of the $N=2$ interface entanglement entropy of \cite{Sakai:2008tt} and \cite{Brehm:2015lja} to the the case of $N\geq 2$ junctions, both for  free boson (\ref{boseEE}) and fermion (\ref{fermEE}) CFTs.  An interesting property of the result is that the both the logarithmically divergent term as well as the constant term only depend on the total transmission coefficient ${\cal T}_i$ into the $i$-th CFT (over which we trace in the entanglement entropy) and the zero mode lattice constant $\text{vol}(\Lambda)$, and thus constitutes the simplest possible generalization of the $N=2$ results. There is an additional term which is regulator dependent and is absent in the $N=2$ case which is independent of the details of the junction.

The most natural extension of these results would be the calculation of the entanglement entropy of CFTs $A\subset\{1,\ldots,N\}$ due to CFTs $B=\bar{A}$. We would expect the entanglement entropy result to change only by
\begin{equation}
\mathcal{T}_i\,\,\longrightarrow\,\,\mathcal{T}_A=\sum_{i\in A}\sum_{j\in B}\mathcal{T}_{ij}
\end{equation}
Most of the calculations of section \ref{sec4} would generalize straightforwardly up to (\ref{bosegaussint}) and (\ref{fermgaussint}), however we would not be able to perform the intermediate Gaussian integrals. Instead, we would need to immediately pass the calculation to the determinant of a block circulant matrix whose larger blocks would have more complicated structure.

It would also be interesting to verify that the Ramond junctions produce the same entanglement entropy as the Neveu-Schwarz junctions, as \cite{Brehm:2015lja} showed explicitly for $N=2$. In addition to the modification of the moding, the form of (\ref{ZKfermint}) would include an additional factor containing Grassmann Gaussian integrals relating to the linearization of the additional quadratic exponent in (\ref{NfermRBstate}). Owing to the somewhat different anticommutation relations between the operators in this additional exponent, these Gaussian integrals have a more complicated structure than those handled in this work. Due to modular invariance, the $K$-sheeted partition function is expected to be
\begin{equation}
Z(K)\sim[\eta(2it)]^{-NK/2}[\theta_2(2it)]^{K(N-2)/2+1}\prod_{k=1}^{K-1}\theta_2(\nu_k|2it)
\end{equation}
which would indeed produce the same entanglement entropy as (\ref{fermEE}). One could also consider interfaces carrying Ramond charge after performing fermion parity projections under the total ${\mathbb{Z}_2}^N$ symmetry, as was done in \cite{Brehm:2015lja} for $N=2$, although it is not clear how easily this could be done for arbitrary $N$.

It may be possible to define a fusion product of junctions, e.g. an $N$-junction and an $N'$-junction fusing in $M$ common CFTs into $(N+N'-2M)$-junctions connecting the remaining CFTs. It might also be interesting to consider if the left/right entanglement entropy calculations of \cite{PandoZayas:2014wsa,Zayas:2016drv,Das:2015oha,Schnitzer:2015gpa} can be extended to D-brane boundary states corresponding to $N$-junctions.

In section \ref{sec2} we have characterized the completely transmissive $N$-junctions as those enforcing twisted permutation gluing conditions. In rational CFTs we could generalize the twisted partition functions of \cite{Petkova:2000ip} to study ``topological" junction operators and their entanglement entropy as in \cite{Gutperle:2015kmw} and \cite{Brehm:2015plf}.

One could also proceed with the type IIB supergravity solutions in \cite{Chiodaroli:2010mv} and calculate the asymmetric 3-junction entanglement entropy holographically as in \cite{Gutperle:2015hcv}. It would be interesting to see if the remarkable holographic agreement in the BPS case between the supergravity calculation and the toy model CFT (i.e. interfaces and junctions of single $c=3/2$ CFTs without reference to the symmetric orbifold) continues to hold for $N=3$. Exploring the case $N=4$ would be more difficult, as there exist D-brane states there that cannot be constructed using successive rotations and T-duality transformations of the elevated $N=3$ D-brane states. Also, the explicit supergravity solutions for $N\geq 4$ have not been found.

\section*{Acknowledgments}
The work reported in this paper is supported in part by the National Science Foundation under grant PHY-16-19926.

\pagebreak

\appendix
\section{CFT conventions}

In this appendix we review the explicit CFT conventions that we use throughout the paper, specifically the free boson and free fermion theories on the cylinder and torus.

For a cylinder of circumference $2\pi$ the action
\begin{equation}
S[\varphi]=\frac{1}{4\pi\alpha'}\int d^2x\,\partial_\mu\varphi\,\partial^\mu\varphi
\end{equation}
describes the compact free boson field $\varphi(x,t)=\varphi(x+2\pi,t)-2\pi wR$, where $w$ is the integer winding number of the boson around the cylinder and $R$ is the compactification radius. The equation of motion is satisfied by
\begin{align}\label{fullmodeexp}
\varphi(z,\bar{z}) =\varphi_0 &- i\paren{\frac{n\alpha'}{2R}+\frac{1}{2}\,wR}\ln z+i\sqrt{\frac{\alpha'}{2}}\sum_{k\neq 0}\frac{1}{k}\,a_k z^{-k}\nonumber\\
&\,\,\,\,\,\,\,\,\,\,-i\paren{\frac{n\alpha'}{2R}-\frac{1}{2}\,wR}\ln\bar{z}+i\sqrt{\frac{\alpha'}{2}}\sum_{k\neq 0}\frac{1}{k}\,\bar{a}_k\bar{z}^{-k}
\end{align}
with holomorphic and anti-holomorphic coordinates given by
\begin{equation}
z = e^{t+ix}\,,\,\,\,\,\,\,\bar{z} = e^{t-ix}
\end{equation}
If we define
\begin{equation}
a_0=\frac{n}{R}\sqrt{\frac{\alpha'}{2}}+\frac{wR}{\sqrt{2\alpha'}}\,,\,\,\,\,\,\bar{a}_0=\frac{n}{R}\sqrt{\frac{\alpha'}{2}}-\frac{wR}{\sqrt{2\alpha'}}
\end{equation}
then the mode expansion (\ref{fullmodeexp}) is brought into the simpler holomorphic and anti-holomorphic expressions
\begin{equation}
i\partial\varphi(z)=\sqrt{\frac{\alpha'}{2}}\sum_{k=-\infty}^\infty a_k z^{-k-1}\,,\,\,\,\,\,i\bar{\partial}\bar{\varphi}(\bar{z})=\sqrt{\frac{\alpha'}{2}}\sum_{k=-\infty}^\infty \bar{a}_k \bar{z}^{-k-1}
\end{equation}
Radial quantization on the complex plane imposes the commutation relations between the bosonic operators (formerly expansion coefficients)
\begin{equation}
[a_n,a_m]=[\bar{a}_n,\bar{a}_m]=n\,\delta_{n+m,0}\,,\,\,\,\,\,[a_n,\bar{a}_m]=0
\end{equation}
The Hamiltonian of this boson (on the torus) is now
\begin{equation}
H=L_0+\bar{L}_0-\frac{1}{12}
\end{equation}
with Virasoro generators given by
\begin{equation}\label{bosenVir}
L_n=\frac{1}{2}\sum_{m=-\infty}^\infty :a_{n-m}a_m:\,,\,\,\,\,\,\,\bar{L}_n=\frac{1}{2}\sum_{m=-\infty}^\infty :\bar{a}_{n-m}\bar{a}_m:
\end{equation}
for $n\neq 0$ and
\begin{equation}\label{bose0Vir}
L_0=\sum_{n=1}^\infty a_{-n}a_n+\frac{1}{2}\,a_0^2\,,\,\,\,\,\,\bar{L}_0=\sum_{n=1}^\infty\bar{a}_{-n}\bar{a}_n+\frac{1}{2}\,\bar{a}_0^2
\end{equation}
The ground state quantum numbers, the momentum and winding number $n$ and $w$, are related to the eigenvalues of the zero mode operators by
\begin{equation}\label{zmeigen}
\sqrt{\frac{1}{2\alpha'}}\paren{a_0+\bar{a}_0}|n,w\rangle=\frac{n}{R}\,|n,w\rangle\,,\,\,\,\,\,\sqrt{\frac{1}{2\alpha'}}\paren{a_0-\bar{a}_0}|n,w\rangle=\frac{mR}{\alpha'}\,|n,w\rangle
\end{equation}
The action of the Hamiltonian on these vacuum states is
\begin{equation}
H\,|n,w\rangle = \paren{\frac{n^2\alpha'}{2R^2}+\frac{w^2R^2}{2\alpha'}-\frac{1}{12}}|n,w\rangle
\end{equation}
With these conventions, the effects of a T-duality transformation are
\begin{equation}\label{Tdualrules}
n\longleftrightarrow w\,,\,\,\,\,\,R\longleftrightarrow\frac{\alpha'}{R}\,,\,\,\,\,\,a_n\longleftrightarrow a_n\,,\,\,\,\,\,\bar{a}_n\longleftrightarrow -\bar{a}_n
\end{equation}

The free Majorana fermion on the cylinder is described by the action
\begin{equation}
S[\psi,\bar{\psi}]=\frac{1}{2\pi\alpha'}\int d^2z\paren{\bar{\psi}\,\partial\bar{\psi}+\psi\,\bar{\partial}\psi}
\end{equation}
where $\psi$ and $\bar{\psi}$ are the component spinors of the Majorana fermion. The equations of motion simply require $\psi(z)$ a holomorphic function and $\bar{\psi}(\bar{z})$ an anti-holomorphic function. These spinors be chosen to be either periodic $\psi(ze^{2\pi i})=\psi(z)$ or anti-periodic $\psi(ze^{2\pi i})=-\psi(z)$. The anti-periodic spinors are said to be in the Neveu-Schwarz sector and have mode expansions
\begin{equation}
i\psi(z)=\sum_{n\in\mathbb{Z}-\tfrac{1}{2}}\psi_n\,z^{-n-1/2}\,\,\,\,\,\,\text{and}\,\,\,\,\,\,i\bar{\psi}(\bar{z})=\sum_{n\in\mathbb{Z}-\tfrac{1}{2}}\bar{\psi}_n\,\bar{z}^{-n-1/2}
\end{equation}
The periodic spinors are said to be in the Ramond sector and have mode expansions
\begin{equation}
i\psi(z)=\sum_{n=-\infty}^\infty\psi_n\,z^{-n-1/2}\,\,\,\,\,\,\text{and}\,\,\,\,\,\,i\bar{\psi}(\bar{z})=\sum_{n=-\infty}^\infty\bar{\psi}_n\,\bar{z}^{-n-1/2}
\end{equation}
In either case, radial quantization on the complex plane imposes anti-commutation relations between the fermionic operators (formerly expansion coefficients)
\begin{equation}\label{fermacoms}
\{\psi_n,\psi_m\}=\{\bar{\psi}_n,\bar{\psi}_m\}=\delta_{n+m,0}\,,\,\,\,\,\,\{\psi_n,\bar{\psi}_m\}=0
\end{equation}
The Hamiltonian of this fermion (on the torus) is now
\begin{equation}
H=L_0+\bar{L}_0-\frac{1}{24}
\end{equation}
with Virasoro generators given by
\begin{equation}
L_n=\frac{1}{2}\sum_{m} \paren{m+\tfrac{1}{2}}:\psi_{n-m}\psi_m:\,,\,\,\,\,\,\,\bar{L}_n=\frac{1}{2}\sum_{m} \paren{m+\tfrac{1}{2}}:\bar{\psi}_{n-m}\bar{\psi}_m:
\end{equation}
for $n\neq 0$, where $m$ is summed over the half-integers or integers for the Neveu-Schwarz or Ramond sectors. For the Neveau-Schwarz sector the $n=0$ generators are
\begin{equation}
L_0=\sum_{n\in\mathbb{N}-\tfrac{1}{2}}n\,\psi_{-n}\psi_n\,,\,\,\,\,\,\bar{L}_0=\sum_{n\in\mathbb{N}-\tfrac{1}{2}}n\,\bar{\psi}_{-n}\bar{\psi}_n
\end{equation}
and for the Ramond sector the $n=0$ generators are
\begin{equation}
L_0=\sum_{n=1}^\infty n\,\psi_{-n}\psi_n+\frac{1}{16}\,,\,\,\,\,\,\bar{L}_0=\sum_{n=1}^\infty n\,\bar{\psi}_{-n}\bar{\psi}_n+\frac{1}{16}
\end{equation}
The action of the Neveu-Schwarz Hamiltonian on the vacuum state is
\begin{equation}
H\,|0\rangle = -\frac{1}{24}\,|0\rangle
\end{equation}
and the action of the Ramond Hamiltonian on the vacuum states is
\begin{equation}
H\,|\pm\rangle = \frac{1}{12}\,|\pm\rangle
\end{equation}
The zero mode operators of the Ramond sector have the action on these vacuum states
\begin{equation}
\psi_0|\pm\rangle = \frac{1}{\sqrt{2}}\,e^{\pm i\pi/4}|\mp\rangle\,,\,\,\,\,\,\bar{\psi}_0|\pm\rangle = \frac{1}{\sqrt{2}}\,e^{\mp i\pi/4}|\mp\rangle
\end{equation}
furnishing a representation of (\ref{fermacoms}) for $n=m=0$.

As a final note, specific values of the coupling $\alpha'$ are often chosen in the literature. In \cite{Sakai:2008tt} and \cite{Brehm:2015lja} the authors use $\alpha'=1/2$. In other works, e.g. \cite{Chiodaroli:2010mv}, $\alpha'=2$ is used.

\section{Special functions}
\subsection{Theta functions and $S$-transformations}\label{thetaS}
The fundamental theta function we use, sometimes called a lattice theta function, is
\begin{equation}
\Theta_\Lambda(\tau)=\sum_{\boldsymbol\lambda\in\Lambda}e^{\pi i\tau|\boldsymbol\lambda|^2}
\end{equation}
Poisson resummation yields the $S$-transformation
\begin{equation}\label{latticeStrans}
\Theta_{\Lambda^*}(-1/\tau)=(-i\tau)^{d/2}\,\text{vol}(\Lambda)\,\Theta_\Lambda(\tau)
\end{equation}
where $\Lambda^*$ is the dual lattice to $\Lambda$, $\text{vol}(\Lambda)$ is the volume of the unit cell, and $d$ is the dimension of the lattice. When a basis of $\Lambda$ is known; that is, when we have a set of $d$ linearly independent vectors $\{\boldsymbol\epsilon_1,\ldots,\boldsymbol\epsilon_d\}$, $\boldsymbol\epsilon_i\in\mathbb{R}^N$, such that
\begin{align}
\Lambda=\left\{\sum_{i=1}^dm^i\boldsymbol\epsilon_i\,\middle|\,\mathbf{m}\in\mathbb{Z}_d\right\}
\end{align}
then $\text{vol}(\Lambda)$ and the basis of $\Lambda^*$ can be computed directly. Let $B$ be the $N\times d$ matrix whose columns are the basis vectors $\boldsymbol\epsilon_i$. In terms of this matrix, the volume of the unit cell is
\begin{equation}
\text{vol}(\Lambda)=\sqrt{\det\paren{B^\text{T}B}}
\end{equation}
and the dual basis is taken from the columns of
\begin{equation}
B^*=B\paren{B^\text{T}B}^{-1}
\end{equation}
As in section 5, sometimes only a set of generators of $\Lambda$ is known; that is, when we have a set of $D>d$ real vectors $\{\boldsymbol\epsilon_1,\cdots,\boldsymbol\epsilon_d,\boldsymbol\delta_1,\cdots,\boldsymbol\delta_{D-d}\}$ such that
\begin{align}
\Lambda=\left\{\sum_{i=1}^dm^i\boldsymbol\epsilon_i+\sum_{j=1}^{D-d}\gamma^j\boldsymbol\delta_j\,\middle|\,\mathbf{m}\in\mathbb{Z}_d,\,\boldsymbol\gamma\in\Gamma\right\}
\end{align}
where the $\boldsymbol\epsilon_i$ are linearly independent and $\Gamma$ is a finite subset of $\mathbb{Z}_{D-d}$ (containing the origin). Additionally we require that $\Gamma$ is chosen such that each point in $\Lambda$ has a unique representation in terms of linear combinations of the above form. This amounts to describing the lattice in terms of a superposition of a finite number of distinct translations of a $d$-dimensional sublattice with a known basis.

In either case the lattice theta function can be expressed in terms of more conventional theta functions. The multi-dimensional theta functions with characteristics (see \cite{olver2010nist} for a wide range of properties) are given by
\begin{equation}
\Theta_d\mathopen{}\begin{bmatrix} \boldsymbol\alpha \\ \boldsymbol\beta \end{bmatrix}\mathclose{}(\mathbf{z}|\Omega)=\sum_{\mathbf{n}\in\mathbb{Z}_d}e^{2\pi i\left[\tfrac{1}{2}(\mathbf{n}+\boldsymbol\alpha)\cdot\Omega(\mathbf{n}+\boldsymbol\alpha) +(\mathbf{n}+\boldsymbol\alpha)\cdot(\mathbf{z}+\boldsymbol\beta)\right]}
\end{equation}
where $\Omega$ is a $d\times d$ matrix. Using Poisson resummation, the action of an $S$-transformation is given by
\begin{equation}
\Theta_d\mathopen{}\begin{bmatrix} -\boldsymbol\beta \\ \boldsymbol\alpha \end{bmatrix}\mathclose{}(\Omega^{-1}\mathbf{z}|-\Omega^{-1})=\sqrt{\det\paren{-i\Omega}}\,e^{-2\pi i\boldsymbol\alpha\cdot\boldsymbol\beta+\pi i\mathbf{z}\cdot\Omega^{-1}\mathbf{z}}\,\Theta_d\mathopen{}\begin{bmatrix} \boldsymbol\alpha \\ \boldsymbol\beta \end{bmatrix}\mathclose{}(\mathbf{z}|\Omega)
\end{equation}
For zero characteristics
$$\Theta_d(\mathbf{z}|\Omega)\equiv\Theta_d\mathopen{}\begin{bmatrix} \mathbf{0} \\ \mathbf{0} \end{bmatrix}\mathclose{}(\mathbf{z}|\Omega)$$
the $S$-transformation is reduced to
\begin{align}
\Theta_d(\Omega^{-1}\mathbf{z}|-\Omega^{-1}) &= \sqrt{\text{det}(-i\Omega)}\,e^{\pi i\mathbf{z}\cdot\Omega^{-1}\mathbf{z}}\,\Theta_d(\mathbf{z}|\Omega)
\end{align}
The zero characteristic theta functions are related to those with nonzero characteristics through
\begin{equation}
\Theta_d\mathopen{}\begin{bmatrix} \boldsymbol\alpha \\ \boldsymbol\beta \end{bmatrix}\mathclose{}(\mathbf{z}|\Omega)=e^{\pi i\paren{\boldsymbol\alpha\cdot\Omega\boldsymbol\alpha+2\boldsymbol\alpha\cdot\paren{\mathbf{z}+\boldsymbol\beta}}}\,\Theta_d(\mathbf{z}+\Omega\boldsymbol\alpha+\boldsymbol\beta|\Omega)
\end{equation}
When a basis is known, the lattice theta function can be simply written
\begin{equation}
\Theta_\Lambda(\tau) = \Theta_d(\tau B^\text{T}B)
\end{equation}
where by standard convention we omit the first argument when $\mathbf{z}=\mathbf{0}$. For the case of a given generating set we instead have
\begin{equation}
\Theta_\Lambda(\tau) = \sum_{\boldsymbol\gamma\in\Gamma}\Theta_d\mathopen{}\begin{bmatrix} e^{\pi i/3}\paren{B_0^\text{T}B_0}^{-1}B_0^\text{T}B_\delta\boldsymbol\gamma \\ \mathbf{0} \end{bmatrix}\mathclose{}(\tau e^{-\pi i/3}B_0^\text{T}B_\delta\boldsymbol\gamma|\tau B_0^\text{T}B_0)
\end{equation}
where $B_0$ is the basis matrix for the lattice $\Lambda_0$ generated by the set $\{\boldsymbol\epsilon_i\}$ alone and $B_\delta$ is the matrix whose columns are the excess generating vectors $\boldsymbol\delta_j$. Setting $\tau=i\varepsilon$ for $\varepsilon\ll 1$ we perform $S$-transformations to obtain
\begin{align}
\Theta_\Lambda(i\varepsilon) &= \frac{\varepsilon^{-d/2}}{\text{vol}(\Lambda_0)}\sum_{\boldsymbol\gamma\in\Gamma}e^{\varepsilon\pi\boldsymbol\gamma\cdot B_\delta^\text{T}B_\delta\boldsymbol\gamma}\\
&\,\,\,\,\,\,\,\,\,\,\,\,\,\,\,\,\,\,\,\,\times\Theta_d\mathopen{}\begin{bmatrix} \mathbf{0} \\ e^{\pi i/3}\paren{B_0^\text{T}B_0}^{-1}B_0^\text{T}B_\delta\boldsymbol\gamma \end{bmatrix}\mathclose{}\paren{e^{-\pi i/3}\paren{B_0^\text{T}B_0}^{-1}B_0^\text{T}B_\delta\boldsymbol\gamma\,\middle|\,\frac{i}{\varepsilon}\paren{B_0^\text{T}B_0}^{-1}}\nonumber\\
&= \frac{|\Gamma|}{\text{vol}(\Lambda_0)}\,\varepsilon^{-d/2}\,\Big(1+\mathcal{O}[\varepsilon]\Big)\Big(1+\mathcal{O}[e^{-\mu/\varepsilon}]\Big)
\end{align}
where $\mu$ is a positive number independent of $\varepsilon$. Comparing this to the leading order behavior of (\ref{latticeStrans}) for $\tau=i\varepsilon$ we obtain
\begin{equation}\label{volfromgen}
\text{vol}(\Lambda)=\frac{\text{vol}(\Lambda_0)}{|\Gamma|}
\end{equation}
From this relationship we can determine the volume of the unit cell of $\Lambda$ from a set of generators.

Lastly, some special consideration is warranted for one-dimensional theta functions. For the case $d=1$ we use a lowercase theta, replace the matrix argument $\Omega$ with a complex variable $\tau$, and define $q=e^{2\pi i\tau}$ for notational simplicity
\begin{align}\theta[\alpha,\beta](z|\tau)\equiv\Theta_1\mathopen{}\begin{bmatrix} \alpha \\ \beta \end{bmatrix}\mathclose{}(z|\tau)
\end{align}
The one-dimensional theta functions can be written in the form of an infinite product
\begin{equation}
\theta[\alpha,\beta](z|\tau) = e^{2\pi i\alpha(z+\beta)}q^{\alpha^2/2}\prod_{n=1}^\infty\paren{1-q^n}\paren{1+q^{n+\alpha-\tfrac{1}{2}}e^{2\pi i(z+\beta)}}\paren{1+q^{n-\alpha-\tfrac{1}{2}}e^{-2\pi i(z+\beta)}}
\end{equation}
such that the usual Jacobi theta functions
\begin{align}
\theta_1(z|\tau) &= \theta[\tfrac{1}{2},\tfrac{1}{2}](z|\tau)\,,\,\,\,\,\,\,\theta_2(z|\tau) = \theta[\tfrac{1}{2},0](z|\tau)\,,\nonumber\\
\theta_3(z|\tau) &= \theta[0,0](z|\tau)\,,\,\,\,\,\,\,\theta_4(z|\tau) = \theta[0,\tfrac{1}{2}](z|\tau)
\end{align}
have sum and product forms
\begin{align}\label{thetaforms}
\theta_1(z|\tau) &= -i\sum_{n\in\mathbb{Z}+\tfrac{1}{2}}(-1)^{n-\tfrac{1}{2}}q^{n^2/2}e^{2\pi inz} = 2\sin(\pi z)\,q^{1/8}\prod_{n=1}^\infty\paren{1-q^n}\paren{1-2q^n\cos(2\pi z)+q^{2n}}\nonumber\\
\theta_2(z|\tau) &= \sum_{n\in\mathbb{Z}+\tfrac{1}{2}}q^{n^2/2}e^{2\pi inz} =  2\cos(\pi z)\,q^{1/8}\prod_{n=1}^\infty\paren{1-q^n}\paren{1+2q^n\cos(2\pi z)+q^{2n}}\nonumber\\
\theta_3(z|\tau) &= \sum_{n=-\infty}^\infty q^{n^2/2}e^{2\pi inz} = \prod_{n=1}^\infty\paren{1-q^n}\prod_{n\in\mathbb{N}-\tfrac{1}{2}}\paren{1+2q^n\cos(2\pi z)+q^{2n}}\\
\theta_4(z|\tau) &= \sum_{n=-\infty}^\infty(-1)^nq^{n^2/2}e^{2\pi inz}  = \prod_{n=1}^\infty\paren{1-q^n}\prod_{n\in\mathbb{N}-\tfrac{1}{2}}\paren{1-2q^n\cos(2\pi z)+q^{2n}}\nonumber
\end{align}
and $S$-transformations given by
\begin{align}
\theta_1(\tfrac{z}{\tau}|-\tfrac{1}{\tau}) &= i\sqrt{-i\tau}\,e^{\pi iz^2/\tau}\,\theta_1(z|\tau)\nonumber\\
\theta_2(\tfrac{z}{\tau}|-\tfrac{1}{\tau}) &= \sqrt{-i\tau}\,e^{\pi iz^2/\tau}\,\theta_4(z|\tau)\nonumber\\
\theta_3(\tfrac{z}{\tau}|-\tfrac{1}{\tau}) &= \sqrt{-i\tau}\,e^{\pi iz^2/\tau}\,\theta_3(z|\tau)\\
\theta_4(\tfrac{z}{\tau}|-\tfrac{1}{\tau}) &= \sqrt{-i\tau}\,e^{\pi iz^2/\tau}\,\theta_2(z|\tau)\nonumber
\end{align}

\subsection{Dedekind eta and related functions}\label{Dedeta}

The Dedekind eta function is
\begin{equation}\label{Dedetadef}
\eta(\tau)=q^{\tfrac{1}{24}}\prod_{n=1}^\infty\paren{1-q^n}=\sum_{n=-\infty}^\infty(-1)^nq^{(6n-1)^2/24}
\end{equation}
and has the modular transformations
\begin{align}
\eta(\tau+1) &= e^{\tfrac{i\pi}{12}}\,\eta(\tau)\\
\eta(-\tfrac{1}{\tau}) &= \sqrt{-i\tau}\,\eta(\tau)
\end{align}
Two related functions 
\begin{align} \prod_{n=1}^\infty\paren{1+q^n}\,\,\,\,\,\,\text{and}\,\,\,\,\,\,\prod_{n\in\mathbb{N}-\tfrac{1}{2}}\paren{1+q^n}
\end{align}
 can be written in terms of the Dedekind eta and other Jacobi theta functions as
\begin{align}
\prod_{n=1}^\infty\paren{1+q^n} &= q^{-\tfrac{1}{24}}\sqrt{\frac{\theta_2(\tau)}{\eta(\tau)}}\\
\prod_{n\in\mathbb{N}-\tfrac{1}{2}}\paren{1+q^n} &= q^{\tfrac{1}{48}}\sqrt{\frac{\theta_3(\tau)}{\eta(\tau)}}
\end{align}

\subsection{Bernoulli polynomials}\label{bernpolys}
The Bernoulli polynomials are explicitly given by
\begin{equation}
b_m(x)=\sum_{n=0}^m\frac{1}{n+1}\sum_{k=0}^n(-1)^k\begin{pmatrix} n \\ k \end{pmatrix}\paren{x+k}^m
\end{equation}
These polynomials are generated by the function
\begin{equation}
\frac{te^{xt}}{e^t-1}=\sum_{m=0}^\infty b_m(x)\,\frac{t^m}{m!}\,\,\,\,\,\,\text{for}\,\,\,|t|<2\pi
\end{equation}
and satisfy the derivative property
\begin{equation}
b_m'(x)=mb_{m-1}(x)
\end{equation}
for $m\geq 1$, and thus the Bernoulli polynomials form an Appell sequence. The values of these polynomials at zero are called the Bernoulli numbers $b_n=b_n(0)$. The first two Bernoulli numbers are
\begin{align}
b_0 &= b_0(1) = 1\\
b_1 &= -b_1(1) = -\tfrac{1}{2}
\end{align}
For $n>1$ we have the following relations
\begin{align}\label{Bernnums}
b_{2n} &= b_{2n}(1)=4n\,(-1)^n\int_0^\infty\frac{t^{2n-1}\,dt}{1-e^{2\pi t}}\\
b_{2n+1} &= b_{2n+1}(1)=0
\end{align}
Combined with these expressions for the Bernoulli polynomials and numbers, the sum identity
\begin{equation}
\sum_{k=1}^{K-1}k^m = \frac{b_{m+1}(K)-b_{m+1}}{m+1}
\end{equation}
can be used to analytically continue functions of the form
\begin{equation}
F(K)=\sum_{k=1}^{K-1}f\paren{\frac{k}{K}}
\end{equation}
where $f(x)$ is analytic at $x=0$ and whose series expansion converges everywhere on the interval $[0,1]$. If $f(x)$ has these properties we can write
\begin{align}\label{Fancon}
F(K) &= \sum_{m=0}^{\infty}\frac{f^{(m)}(0)}{m!\,K^m}\sum_{k=1}^{K-1}k^m\nonumber\\
&= \sum_{m=0}^{\infty}\frac{f^{(m)}(0)}{(m+1)!\,K^m}\left[b_{m+1}(K)-b_{m+1}\right]
\end{align}
so that in the last line of the above $F(K)$ is now explicitly an analytic function of $K$. More so than $F(K)$ we are interested in
\begin{equation}
F'(K)=-\sum_{m=1}^\infty\frac{mf^{(m)}(0)}{(m+1)!\,K^{m+1}}\left[b_{m+1}(K)-b_{m+1}\right]+\sum_{m=0}^\infty\frac{f^{(m)}(0)}{m!\,K^m}\,b_m(K)
\end{equation}
and
\begin{align}\label{Fanconfinal}
F'(1) &= \sum_{m=0}^\infty\frac{f^{(m)}(0)}{m!}\,b_m(1)\nonumber\\
&= f(0)+\frac{1}{2}\,f'(0)+2\sum_{m=1}^\infty\frac{f^{(2m)}(0)}{(2m-1)!}\,(-1)^m\int_0^\infty\frac{t^{2m-1}\,dt}{1-e^{2\pi t}}\nonumber\\
&= f(0)+\frac{1}{2}\,f'(0)+\int_0^\infty\frac{if'(it)-if'(-it)}{1-e^{2\pi t}}\,dt
\end{align}
In \cite{Sakai:2008tt} and \cite{Brehm:2015lja} (\ref{Fanconfinal}) was calculated for
\begin{equation}
f_\text{bos}(x)=\frac{1}{2\pi}\,\text{arccos}^2(s\sin\pi x)\,\,\,\,\,\,\,\,\,\,\text{and}\,\,\,\,\,\,\,\,\,\,f_\text{ferm}(x)=\frac{1}{2\pi}\,\text{arcsin}^2(s\sin\pi x) 
\end{equation}
to obtain
\begin{equation}\label{anconresult}
\varphi'(1)=\frac{\pi}{2}\,\sigma(s)-\frac{\pi}{8}\,\,\,\,\,\,\,\,\,\,\text{and}\,\,\,\,\,\,\,\,\,\,\vartheta'(1)=\frac{\pi}{4}\,s-\frac{\pi}{2}\,\sigma(s)
\end{equation}
where $\sigma(s)$ is a complicated function containing dilogarithms
\begin{equation}\label{sigmadef}
\sigma(s)=\frac{1}{6}+\frac{s}{3}+\frac{1}{\pi^2}\left[(s+1)\log(s+1)\log s+(s-1)\,\text{Li}_2(1-s)+(s+1)\,\text{Li}_2(-s)\right]
\end{equation}

\section{Intermediate gaussian integrals}

\subsection{Bosonic integrals}\label{boseintapp}

In the following we repeatedly use the one-dimensional complex Gaussian integral
\begin{equation}\label{bose1dint}
\int_{-\infty}^\infty dz\,d\bar{z}\,\,e^{\,az\bar{z}+bz+c\bar{z}}=-\frac{1}{a}\,e^{-bc/a}
\end{equation}
in order to integrate out all of the dependence on the $i$-th integration variables in (\ref{bosegaussint}). This will involve isolating linear factors of these variables in the exponents of (\ref{bosegaussint}) in order to combine them via (\ref{bose1dint}). We show some of the details of this process below.\\

Focusing on the $z_{ni}^{(k)}$, $\bar{z}_{ni}^{(k)}$ integral for an arbitrary fixed $k$, the linear terms in the exponents of (\ref{bosegaussint}) are rewritten as
\begin{align} q^n\sum_l\paren{S_{il}z^{(k)}_{ni}w^{(k)}_{nl}+S_{li}\bar{z}^{(k)}_{nl}\bar{w}^{(k)}_{ni}} = \paren{q^n\sum_jS_{ij}w_{nj}^{(k)}}z^{(k)}_{ni}+\paren{q^nS_{ii}\bar{w}_{ni}^{(k)}}\bar{z}^{(k)}_{ni}+q^n\sum_{j\neq i}S_{ji}\bar{z}^{(k)}_{nj}\bar{w}_{ni}^{(k)}\nonumber\\
\end{align}
\begin{align}
q^n\sum_{j\neq i}\sum_l\paren{S_{jl}z^{(k)}_{nj}w^{(k-1)}_{nl} +S_{lj}\bar{z}^{(k)}_{nl}\bar{w}^{(k-1)}_{nj}} &= q^n\sum_{j\neq i}\sum_lS_{jl}z^{(k)}_{nj}w^{(k-1)}_{nl}+\paren{q^n\sum_{j\neq i}S_{ij}\bar{w}_{nj}^{(k-1)}}\bar{z}_{ni}^{(k)}\nonumber \\
&\,\,\,\,\,\,+q^n\sum_{j,l\neq i}S_{lj}\bar{z}^{(k)}_{nl}\bar{w}^{(k-1)}_{nj}
\end{align}
in order to isolate the $z_{ni}^{(k)}$ and $\bar{z}_{ni}^{(k)}$ factors. Applying (\ref{bose1dint}) to all the $z_{ni}^{(k)}$, $\bar{z}_{ni}^{(k)}$ integrals then yields the new exponential terms
\begin{align}
&\paren{q^n\sum_jS_{ij}w_{nj}^{(k)}}\paren{q^nS_{ii}\bar{w}_{ni}^{(k)}+q^n\sum_{j\neq i}S_{ij}\bar{w}^{(k-1)}_{nj}}\nonumber \\
&\,\,\,\,\,\,\,\,\,\,\,\,\,\,\,\,\,\,\,\,= q^{2n}S_{ii}^2w^{(k)}_{ni}\bar{w}_{ni}^{(k)}+\paren{q^{2n}S_{ii}\sum_{j\neq i}S_{ij}w_{nj}^{(k)}}\bar{w}_{ni}^{(k)}\nonumber\\
&\,\,\,\,\,\,\,\,\,\,\,\,\,\,\,\,\,\,\,\,\,\,\,\,\,\,\,\,\,\,+\paren{q^{2n}S_{ii}\sum_{j\neq i}S_{ij}\bar{w}_{nj}^{(k-1)}}w^{(k)}_{ni}+q^{2n}\sum_{j,l\neq i}S_{ij}S_{jl}w^{(k)}_{nj}\bar{w}^{(k-1)}_{nl}
\end{align}
where we have now isolated the $w_{ni}^{(k)}$ and $\bar{w}_{ni}^{(k)}$ factors for the next round of integration.

Focusing now on the $w_{ni}^{(k)}$, $\bar{w}_{ni}^{(k)}$ integral for an arbitrary fixed $k$, the quadratic term of the exponent is now $-w^{(k)}_{ni}\bar{w}^{(k)}_{ni}/D_n$ after the $z_{ni}$, $\bar{z}_{ni}$ integration, where $D_n=(1-q^{2n}S_{ii}^2)^{-1}$. The remaining linear terms in the exponent are the above linear terms above in addition to those that spectated the $z_{ni}$, $\bar{z}_{ni}$ integration
\begin{align}
\paren{q^n\sum_{j\neq i}S_{ji}\bar{z}^{(k)}_{nj}}\bar{w}^{(k)}_{ni}+\paren{q^n\sum_{j\neq i}S_{ji}z^{(k+1)}_{nj}}w^{(k)}_{ni}+q^n\sum_{j,l\neq i}S_{jl}\paren{z^{(k+1)}_{nj}w^{(k)}_{nl}+\bar{z}^{(k+1)}_{nj}\bar{w}^{(k)}_{nl}}
\end{align}
so that applying (\ref{bose1dint}) to all the $w_{ni}^{(k)}$, $\bar{w}_{ni}^{(k)}$ integrals then yields the new terms
\begin{align}
&D_n\paren{q^n\sum_{j\neq i}S_{ji}z^{(k+1)}_{nj}+q^{2n}S_{ii}\sum_{j\neq i}S_{ij}\bar{w}^{(k-1)}_{nj}}\paren{q^n\sum_{j\neq i}S_{ji}\bar{z}^{(k)}_{nj}+q^{2n}S_{ii}\sum_{j\neq i}S_{ij}w^{(k)}_{nj}}
\end{align}
At this point there are no linear terms remaining that mix variables with the same value of $k$. Once the above terms are simplified and all indices shifted so that $k$ and $k+1$ are the only indices that appear, we recover (\ref{intreduce}) and (\ref{lintermreduce}).

\subsection{Fermionic integrals}\label{fermintapp}

In the following we repeatedly use the one-dimensional complex Grassmann Gaussian integral
\begin{equation}\label{ferm1dint}
-\int d\eta\,d\bar{\eta}\,\,e^{\,a\eta\bar{\eta}+\beta\eta+\bar{\eta}\gamma}=a\,e^{\,\beta\gamma/a}
\end{equation}
for constant $a$ and Grassmann-valued $\beta$ and $\gamma$, in order to integrate out all of the dependence on the $i$-th integration variables in (\ref{fermgaussint}). This will involve isolating linear factors of these variables in the exponents of (\ref{fermgaussint}) in order to combine them via (\ref{ferm1dint}). We show some of the details of this process below.

Focusing on the $\eta_{ni}^{(k)}$, $\bar{\eta}_{ni}^{(k)}$ integrals for an arbitrary fixed $k$, the linear terms in the exponents of (\ref{fermgaussint}) are rewritten as
\begin{align} &iq^n\sum_j\Big(S_{ij}\eta^{(k)}_{ni}\chi^{(k)}_{nj} +S_{ji}\bar{\eta}^{(k)}_{nj}\bar{\chi}^{(k)}_{ni}\Big)\nonumber \\
 =& \paren{-iq^n\sum_jS_{ij}\chi_{nj}^{(k)}}\eta^{(k)}_{ni}+\bar{\eta}^{(k)}_{ni}\paren{iq^nS_{ii}\bar{\chi}_{ni}^{(k)}}+iq^n\sum_{j\neq i}S_{ji}\bar{\eta}^{(k)}_{nj}\bar{\chi}_{ni}^{(k)}\end{align}
and
\begin{align}
iq^n\sum_l\sum_{j\neq i}\paren{S_{jl}\eta^{(k)}_{nj}\chi^{(k-1)}_{nl} +S_{lj}\bar{\eta}^{(k)}_{nl}\bar{\chi}^{(k-1)}_{nj}} &= iq^n\sum_l\sum_{j\neq i}S_{jl}\eta^{(k)}_{nj}\chi^{(k-1)}_{nl}+\bar{\eta}_{ni}^{(k)}\paren{iq^n\sum_{j\neq i}S_{ij}\bar{\chi}_{nj}^{(k-1)}}\nonumber\\
&\,\,\,\,\,\,+iq^n\sum_{l\neq i}\sum_{j\neq i}S_{lj}\bar{\eta}^{(k)}_{nl}\bar{\chi}^{(k-1)}_{nj}
\end{align}
in order to isolate the $\eta_{ni}^{(k)}$ and $\bar{\eta}_{ni}^{(k)}$ factors. Applying (\ref{ferm1dint}) to all the $\eta_{ni}^{(k)}$, $\bar{\eta}_{ni}^{(k)}$ integrals then yields the new terms
\begin{align}
&\paren{-iq^n\sum_jS_{ij}\chi_{nj}^{(k)}}\paren{iq^nS_{ii}\bar{\chi}_{ni}^{(k)}+iq^n\sum_{j\neq i}S_{ij}\bar{\chi}^{(k-1)}_{nj}}\nonumber \\
&=q^{2n}S_{ii}^2\chi^{(k)}_{ni}\bar{\chi}_{ni}^{(k)}+\bar{\chi}_{ni}^{(k)}\paren{-q^{2n}S_{ii}\sum_{j\neq i}S_{ij}\chi_{nj}^{(k)}}+\paren{-q^{2n}S_{ii}\sum_{j\neq i}S_{ij}\bar{\chi}_{nj}^{(k-1)}}\chi^{(k)}_{ni}\nonumber \\
&\,\,\,\,\,\,\,\,\,\,+q^{2n}\sum_{l\neq i}\sum_{j\neq i}S_{ij}S_{il}\chi^{(k)}_{nj}\bar{\chi}^{(k-1)}_{nl}
\end{align}
where we have now isolated the $\chi_{ni}^{(k)}$ and $\bar{\chi}_{ni}^{(k)}$ factors for the next round of integration.

Focusing now on the $\chi_{ni}^{(k)}$, $\bar{\chi}_{ni}^{(k)}$ integral for a arbitrary fixed $k$, the quadratic term of the exponent is now $\chi^{(k)}_{ni}\bar{\chi}^{(k)}_{ni}/D_n$ after the $\eta_{ni}$, $\bar{\eta}_{ni}$ integration, where $D_n=(1+q^{2n}S_{ii}^2)^{-1}$. The remaining linear terms in the exponent are the linear terms above in addition to those that spectated the $\eta_{ni}$, $\bar{\eta}_{ni}$ integration
\begin{align}
\bar{\chi}^{(k)}_{ni}\paren{-iq^n\sum_{j\neq i}S_{ji}\bar{\eta}^{(k)}_{nj}}+\paren{iq^n\sum_{j\neq i}S_{ji}\eta^{(k+1)}_{nj}}\chi^{(k)}_{ni}+iq^n\sum_{l\neq i}\sum_{j\neq i}S_{jl}\paren{\eta^{(k+1)}_{nj}\chi^{(k)}_{nl} +\bar{\eta}^{(k+1)}_{nj}\bar{\chi}^{(k)}_{nl}}\nonumber \\
\end{align}
so that applying (\ref{ferm1dint}) to all the $\chi_{ni}^{(k)}$, $\bar{\chi}_{ni}^{(k)}$ integrals then yields the new terms
\begin{align}
&D_n\paren{iq^n\sum_{j\neq i}S_{ji}\eta^{(k+1)}_{nj}-q^{2n}S_{ii}\sum_{j\neq i}S_{ij}\bar{\chi}^{(k-1)}_{nj}}\paren{-iq^n\sum_{j\neq i}S_{ji}\bar{\eta}^{(k)}_{nj}-q^{2n}S_{ii}\sum_{j\neq i}S_{ij}\chi^{(k)}_{nj}}\end{align}
At this point there are no linear terms remaining that mix variables with the same value of $k$. Once the above terms are simplified and all indices shifted so that $k$ and $k+1$ are the only indices that appear, we recover (\ref{fermgaussintred}) and (\ref{fermgaussintexp}).

\section{Calculation of determinants}\label{detcalcs}
In the determinant calculations there are two special forms of (equal-sized and square) block matrices that we encounter, those of the block circulant form
\begin{equation}
\mathcal{M}_n = \begin{pmatrix} M_0 & M_{n-1} & M_{n-2} & \cdots & M_2 & M_1 \\ M_1 & M_0 & M_{n-1} & \cdots & M_3 & M_2 \\ M_2 & M_1 & M_0 & \cdots & M_4 & M_3 \\ \vdots & \vdots & \vdots & \ddots & \vdots & \vdots \\ M_{n-2} & M_{n-3} & M_{n-4} & \cdots & M_0 & M_{n-1} \\ M_{n-1} & M_{n-2} & M_{n-3} & \cdots & M_1 & M_0 \end{pmatrix}
\end{equation}
and $2\times 2$ block matrices. The determinant of the block circulant matrix was shown in \cite{Tee05eigenvectorsof} to be
\begin{equation}\label{cirdet}
\det\,\mathcal{M}_n=\prod_{k=1}^n\det\paren{\,\sum_{j=0}^{n-1}e^{2jk\pi i/n}M_j}
\end{equation}
This result is remarkable as (\ref{cirdet}) is of the same form regardless of the size of the matrices $M_j$, including when they reduce to scalars. In general, determinants of block matrices only exhibit similar behavior either when all block entries commute \cite{Silvester:10.2307/3620776}, or when certain blocks are invertible and commute. Consider the $2\times 2$ block matrix
\begin{equation}
\begin{pmatrix} A & B \\ C & D \end{pmatrix}
\end{equation}
with $A$, $B$, $C$, and $D$ all square matrices of the same dimensions. If $A$ is invertible, then the decomposition
\begin{equation}
\begin{pmatrix} A & B \\ C & D \end{pmatrix}=\begin{pmatrix} A & 0 \\ C & 1 \end{pmatrix}\begin{pmatrix} 1 & A^{-1}B \\ 0 & D-CA^{-1}B \end{pmatrix}
\end{equation}
leads to the determinant equation
\begin{equation}
\det\begin{pmatrix} A & B \\ C & D \end{pmatrix}=\det\,A\,\det\paren{D-CA^{-1}B}
\end{equation}
If we also have that $[A,C]=0$ then the determinant reduces to
\begin{equation}
\det\begin{pmatrix} A & B \\ C & D \end{pmatrix}=\det\paren{AD-CB}
\end{equation}
while if $[A,B]=0$ the determinant becomes
\begin{equation}\label{22bmdet}
\det\begin{pmatrix} A & B \\ C & D \end{pmatrix}=\det\paren{DA-CB}
\end{equation}
Similar results holds if $D$ is invertible and $[C,D]=0$ or $[B,D]=0$.
\subsection{Bosonic determinant}
Beginning with the matrix defined in (\ref{boseMK}), (\ref{boseC}), and (\ref{boseXYZ}) we apply (\ref{cirdet}) to obtain
\begin{align}
\det\,M_K &= \prod_{k=1}^K\det\paren{1_{4N-4}+e^{2\pi ik/K}C+e^{-2\pi ik/K}C^\text{T}\,}\nonumber\\
&=\prod_{k=1}^K\det\begin{pmatrix} 1_{2N-2}+X\otimes U_k & e^{2\pi ik/K}Y\otimes\sigma^3 \\ e^{-2\pi ik/K}Y^\text{T}\otimes\sigma^3 & 1_{2N-2}+Z\otimes U_k \end{pmatrix}
\end{align}
where
\begin{equation}\label{Ukdef}
U_k = \cos(2\pi k/K)\,1_2+i\sin (2\pi k/K)\,\sigma^2 = \exp\paren{2\pi ik\sigma^2/K}
\end{equation}
In order to analyze the structure of the $2\times 2$ block matrices above, we calculate a few properties of the blocks (\ref{boseXYZ})
\begin{equation}\label{startXYZprops}
\text{Tr}\,[X]=\text{Tr}\,[Z]=-q^{2n}D_n\paren{1-S_{ii}^2}
\end{equation}
\begin{equation}\label{xzgenproj}
X^2=-q^{2n}D_n\paren{1-S_{ii}^2}X\,,\,\,\,\,\,\,Z^2=-q^{2n}D_n\paren{1-S_{ii}^2}Z
\end{equation}
\begin{align}
(XY)_{jl} = (YZ)_{jl} &= -q^{3n}D_n^2S_{ii}\paren{1-q^{2n}}S_{ji}S_{il}\\
YY^\text{T} &= q^{2n}1_{N-1}+D_n\paren{1-q^{4n}S_{ii}^2}X\label{yytran}\\
Y^\text{T}Y &= q^{2n}1_{N-1}+D_n\paren{1-q^{4n}S_{ii}^2}Z\label{ytrany}
\end{align}
From (\ref{xzgenproj}) we see that $\det\,X=\det\,Z=0$, and hence $X$ and $Z$ are not invertible. However, employing the matrix logarithm, the Mercator series, and the geometric series we find
\begin{align}
\det\paren{1_{2N-2}+X\otimes U_k} &= \exp\paren{-\sum_{m=1}^\infty\frac{(-1)^m}{m}\,\text{Tr}[X^m]\,\text{Tr}[U_k^m]}\nonumber\\
&= \exp\paren{-\sum_{m=1}^\infty\frac{1}{m}\,\paren{q^{2n}D_n(1-S_{ii}^2)}^m\paren{e^{2\pi imk/K}+e^{-2\pi imk/K}}}\nonumber\\
&= 1-2q^{2n}D_n(1-S_{ii}^2)\cos(2\pi k/K)+q^{4n}D_n^2(1-S_{ii}^2)^2\\
&=\det\paren{1_{2N-2}+Z\otimes U_k}\nonumber
\end{align}
Thus $1_{2N-2}+X\otimes U_k$ and $1_{2N-2}+Z\otimes U_k$ are both invertible. A very similar determinant calculation using (\ref{yytran}) and (\ref{ytrany}) shows that $\det\,Y\neq 0$ and hence $Y$ is invertible. At this point we make the decomposition
\begin{align}\label{Mkdecomp}
&\begin{pmatrix} 1_{2N-2}+X\otimes U_k & e^{2\pi ik/K}Y\otimes\sigma^3 \\ e^{-2\pi ik/K}Y^\text{T}\otimes\sigma^3 & 1_{2N-2}+Z\otimes U_k \end{pmatrix}=\nonumber\\
&\,\,\,\,\,\,\,\,\,\,\,\,\,\,\,\,\,\,\,\,\begin{pmatrix} e^{2\pi ik/K}B_k & e^{2\pi ik/K}Y\otimes\sigma^3 \\ 1_{2N-2} & 1_{2N-2}+Z\otimes U_k \end{pmatrix}\begin{pmatrix} e^{2\pi ik/K}Y\otimes\sigma^3A_k^{-1} & 0 \\ 1_{N-1}\otimes\paren{1_2-q^{2n}A_k^{-1}} & 1_{2N-2} \end{pmatrix}^{-1}
\end{align}
with matrices
\begin{equation}\label{amtxdef}
A_k=q^{2n}1_2-D_n\paren{1-q^{4n}S_{ii}^2}U_k^{-1}
\end{equation}
and
\begin{equation}\label{bmtxdef}
B_k=Y\otimes\sigma^3\paren{1_2+(1-q^{2n})A_k^{-1}}+XY\otimes U_k\sigma^3A_k^{-1}
\end{equation}
Now using (\ref{Mkdecomp}) and (\ref{22bmdet}), the determinant can be reduced to
\begin{align}
\det\,M_K &= \prod_{k=1}^K\det\left[B_k\paren{Y^{-1}\otimes A_k\sigma^3}+B_k\paren{Z\otimes U_k}\paren{Y^{-1}\otimes A_k\sigma^3}-1_{N-1}\otimes\sigma^3A_k\sigma^3\right]\nonumber\\
&= \prod_{k=1}^K\paren{1-q^{2n}}^{2N-2}\det\paren{1_{2N-2}+\frac{2\cos(2\pi k/K)-q^{2n}-1}{1-q^{2n}}\,X\otimes 1_2}\nonumber\\
&= D_n^{2K}\prod_{k=1}^K\paren{1-q^{2n}}^{2N-4}\left[1-2\paren{S_{ii}^2+\paren{1-S_{ii}^2}\cos(2\pi k/K)}q^{2n}+q^{4n}\right]^2
\end{align}

\subsection{Fermionic determinant}
In this case the block entries (\ref{boseXYZ}) and their properties in (\ref{startXYZprops}) through (\ref{ytrany}) are modified by $q^n\rightarrow -iq^n$. We proceed in a similar manner to the previous section, where now
\begin{align}
\det\,M_K &= \prod_{k=1}^K\det\paren{1_{2N-2}\otimes\sigma^2+e^{2\pi ik/K}C-e^{-2\pi ik/K}C^\text{T}\,}\nonumber\\
&=\prod_{k=1}^K\det\begin{pmatrix} 1_{N-1}\otimes\sigma^2+X\otimes U_k\sigma^2 & e^{2\pi ik/K}Y\otimes\sigma^3 \\ -e^{-2\pi ik/K}{Y}^\text{T}\otimes\sigma^3 & 1_{N-1}\otimes\sigma^2+Z\otimes U_k\sigma^2 \end{pmatrix}\nonumber\\
&=\paren{-1}^{2(N-1)K}\prod_{k=1}^K\det\begin{pmatrix} 1_{2N-2}+X\otimes U_k & -ie^{2\pi ik/K}Y\otimes\sigma^1 \\ ie^{-2\pi ik/K}{Y}^\text{T}\otimes\sigma^1 & 1_{2N-2}+Z\otimes U_k \end{pmatrix}
\end{align}
with $U_k$ as in (\ref{Ukdef}). Making the decomposition
\begin{align}\label{Mkdecomp2}
&\begin{pmatrix} 1_{2N-2}+X\otimes U_k & -ie^{2\pi ik/K}Y\otimes\sigma^1 \\ ie^{-2\pi ik/K}{Y}^\text{T}\otimes\sigma^1 & 1_{2N-2}+Z\otimes U_k \end{pmatrix}=\nonumber\\
&\,\,\,\,\,\,\,\,\,\,\,\,\,\,\,\,\,\,\,\,\begin{pmatrix} -ie^{2\pi ik/K}B_k & -ie^{2\pi ik/K}Y\otimes\sigma^1 \\ 1_{2N-2} & 1_{2N-2}+Z\otimes U_k \end{pmatrix}\begin{pmatrix} -ie^{2\pi ik/K}Y\otimes\sigma^1A_k^{-1} & 0 \\ 1_{N-1}\otimes\paren{1_2+q^{2n}A_k^{-1}} & 1_{2N-2} \end{pmatrix}^{-1}
\end{align}
with matrices
\begin{equation}
A_k=-q^{2n}1_2-D_n\paren{1-q^{4n}S_{ii}^2}U_k^{-1}
\end{equation}
and
\begin{equation}
B_k=Y\otimes\sigma^1\paren{1_2+(1+q^{2n})A_k^{-1}}+XY\otimes U_k\sigma^1A_k^{-1}
\end{equation}
we use (\ref{Mkdecomp2}) and (\ref{22bmdet}) to reduce the determinants to
\begin{align}
\det\,M_K &= \paren{-1}^{2(N-1)K}\prod_{k=1}^K\paren{1+q^{2n}}^{2N-2}\det\paren{1_{2N-2}+\frac{2\cos(2\pi k/K)+q^{2n}-1}{1+q^{2n}}\,X\otimes 1_2}\nonumber\\
&= D_n^{2K}\paren{-1}^{2(N-1)K}\prod_{k=1}^K\paren{1+q^{2n}}^{2N-4}\left[1+2\paren{S_{ii}^2+\paren{1-S_{ii}^2}\cos(2\pi k/K)}q^{2n}+q^{4n}\right]^2
\end{align}

\bibliographystyle{utphys}
\providecommand{\href}[2]{#2}\begingroup\raggedright\endgroup

\end{document}